\documentclass[12pt]{article}

\usepackage{cite}

\newcommand{\om}{\omega}
\newcommand{\al}{\alpha}
\newcommand{\ep}{\epsilon}

\newcommand{\la}{\lambda}

\newcommand{\hmu}{\hat{\mu}}

\newcommand{\NS}{\mbox{NS}}
\newcommand{\tNS}{\widetilde{\mbox{NS}}}
\newcommand{\R}{\mbox{R}}
\newcommand{\tR}{\widetilde{\mbox{R}}}
\newcommand{\sNS}{\msc{NS}}
\newcommand{\stNS}{\widetilde{\msc{NS}}}
\newcommand{\sR}{\msc{R}}
\newcommand{\stR}{\widetilde{\msc{R}}}

\newcommand{\lb}{\lbrack}
\newcommand{\rb}{\rbrack}

\newcommand{\msc}[1]{\mbox{\scriptsize #1}}
\newcommand{\dsp}{\displaystyle}

\newcommand{\br}{\mbox{{\bf R}}}
\newcommand{\bz}{\mbox{{\bf Z}}}

\newcommand{\bs}{\mbox{{\bf S}}}

\newcommand{\bsz}{\msc{{\bf Z}}}

\newcommand{\cN}{{\cal N}}

\newcommand{\cC}{{\cal C}}
\newcommand{\cQ}{{\cal Q}}

\newcommand{\tL}{\tilde{L}}
\newcommand{\tJ}{\tilde{J}}
\newcommand{\tG}{\tilde{G}}
\newcommand{\tF}{\tilde{F}}

\newcommand{\ket}[1]{{|#1\rangle}}
\newcommand{\bra}[1]{{\langle#1|}}
\newcommand{\dket}[1]{{\left.\left|#1\right\rangle\right\rangle}}
\newcommand{\dbra}[1]{{\left\langle\left\langle#1\right|\right.}}
\newcommand{\dketm}[1]{{\left.\left|#1\right\rangle\right
 \rangle_{\msc{\bf M}}}}
\newcommand{\dbram}[1]{{{}_{\msc{\bf M}}
\left\langle\left\langle#1\right|\right.}}
\newcommand{\ketA}[1]{{|#1\rangle}_{\msc{A}}}
\newcommand{\braA}[1]{{}_{\msc{A}}{\langle#1|}}

\newcommand{\Th}[2]{\Theta_{#1,#2}}
\renewcommand{\th}{{\theta}}

\newcommand{\ch}[2]{\mbox{ch}^{#1}_{#2}}

\newcommand{\Ch}[2]{\mbox{Ch}^{#1}_{#2}}

\newcommand{\chm}[1]{\mbox{ch}^{#1}_{\msc{\bf M}}}
\newcommand{\chg}[1]{\mbox{ch}^{#1}_{\msc{\bf G}}}
\newcommand{\chim}{\chi_{\msc{\bf M}}}
\newcommand{\chig}{\chi_{\msc{\bf G}}}

\newcommand{\Chm}[1]{\mbox{Ch}^{#1}_{\msc{\bf M}}}
\newcommand{\Chg}[1]{\mbox{Ch}^{#1}_{\msc{\bf G}}}

\newcommand{\mod}{\mbox{mod}}

\newcommand{\nn}{\nonumber\\}

\newcommand {\eqn}[1]{(\ref{#1})}

\makeatletter
\@addtoreset{equation}{section}
\def\theequation{\thesection.\arabic{equation}}
\makeatother

\begin{document}
\vskip 7mm

\begin{titlepage}
 \
 \renewcommand{\thefootnote}{\fnsymbol{footnote}}
 \font\csc=cmcsc10 scaled\magstep1
 {\baselineskip=14pt
 \rightline{
 \vbox{\hbox{hep-th/0311141}
       \hbox{UT-03-37}
       }}}

 \baselineskip=20pt
\vskip 2cm
 
\begin{center}
 \centerline{\huge  Modular Bootstrap }

\vskip 5mm

{\huge for Boundary $\cN=2$ Liouville Theory} 

 \vskip 2cm
\noindent{ \large Tohru Eguchi  and Yuji Sugawara} \\
{\sf eguchi@hep-th.phys.s.u-tokyo.ac.jp~,~
sugawara@hep-th.phys.s.u-tokyo.ac.jp}
\bigskip

 \vskip .6 truecm
 {\baselineskip=15pt
 {\it Department of Physics,  Faculty of Science, \\
  University of Tokyo \\
  Hongo 7-3-1, Bunkyo-ku, Tokyo 113-0033, Japan}
 }
\end{center}

\bigskip 

\bigskip

\begin{abstract}

We study the boundary $\cN=2$ Liouville theory based on 
the ``modular bootstrap'' approach. As fundamental conformal blocks
we introduce the ``extended characters'' that are  defined as 
the proper sums over spectral flows of irreducible characters of the $\cN=2$ 
superconformal algebra (SCA) 
and clarify their modular transformation properties 
in models with rational central
charges. We then try to classify the Cardy states describing 
consistent $D$-branes based on the modular data. 
We construct the analogues of ZZ-branes \cite{ZZ},
localized at the strong coupling region,
and the FZZT-branes \cite{FZZ,Teschner}, which extend along the
Liouville direction. The former is shown to play important roles
to describe the BPS $D$-branes wrapped around vanishing cycles in
deformed Calabi-Yau singularities, reproducing the correct 
intersection numbers of vanishing cycles. We also discuss the
non-BPS $D$-branes in 2d type 0 (and type II) string vacua composed of the $\cN=2$ Liouville 
with $\hat{c}(\equiv c/3)=5$. 
Unstable D0-branes are found as the ZZ-brane analogues
mentioned above, and the FZZT-brane analogues are stable due to 
the existence of mass gap despite the lack of GSO projection.

\end{abstract}

\vfill

\setcounter{footnote}{0}
\renewcommand{\thefootnote}{\arabic{footnote}}
\end{titlepage}
\baselineskip 18pt

\section{Introduction}

~

Study of $D$-branes in the $\cN=2$ super-Liouville theory \cite{KutS}
is one of the subjects of great importance by several reasons.  
First of all, irrational boundary (super)conformal field 
theories still have many open problems and subtleties,
and the $\cN=2$ boundary Liouville theory will yield  
a good chance to grasp the essence of the issues. 
Even though the boundary (bosonic) Liouville theory 
has been firmly understood \cite{FZZ,Teschner,ZZ} 
and the $\cN=1$ boundary Liouville can be 
investigated in a parallel way \cite{FH,ARS}, 
the $\cN=2$ Liouville is still hard to analyse and challenging,
since it has an extra $U(1)$-sector coupled with 
the Liouville field.

The second reason lies in the $\cN=2$ SCFT descriptions of 
singular Calabi-Yau compactifications or the 
suitable NS5 configurations in the T-dual picture 
\cite{GV,OV,ABKS,GKP,GK,Pelc,ES1,Mizoguchi,Yamaguchi,NN,HK2}.
Even under the $g_s\,\rightarrow\,0$ limit, the divergence of dilaton 
at the singular locus makes the relevant conformal system non-trivial.
The $\cN=2$ Liouville sector plays a central role in these 
superconformal system and turning on the Liouville potential
deforms the singularity. The boundary state analysis in the $\cN=2$
Liouville theory is thus important to describe the BPS $D$-branes 
wrapped around the vanishing cycles. Earlier attempts at this subject
have been given in {\em e.g.}  \cite{Lerche,LLS,ES2}.

The third one originates from recent studies on the 
proposed dualities between $d\leq 2$ string vacua with 
matrix models from the viewpoints of  unstable $D$-branes
\cite{MV,KMS,MTV,Martinec,AKK,TT,DKKMMS,GIR,GutK,Sen,MMV,Kapustin,GKPS,KStro,
DRSVW,KMS2,DD,Alex,GKap,Mukhi,BH}. 
The $\cN=2$ Liouville shares the same field content
as the 2d type 0 string vacua constructed from the $c=3/2$ 
superconformal matter coupled with $\cN=1$ Liouville studied in
\cite{TT,DKKMMS,Kapustin,DRSVW,GKap,BH}. However, they have different 
world-sheet interactions, 
and hence it is an interesting problem to investigate whether we can 
discover the similar dual description by matrix models
for the $\cN=2$ Liouville. Several analyses and conjectures are 
found in the recent works \cite{MMV,GKPS}.

The standard method to determine the proper boundary states in the Liouville 
theory is to solve the bootstrap equation for the disc one-point function
\cite{FZZ,ZZ}. As we mentioned above, however, this prescription 
is likely  to be difficult for the $\cN=2$ Liouville case.
We shall hence take a different route that is easier 
from a technical point of view: we use the ``modular bootstrap''
based on the modular transformation properties of character formulas. 
Although we cannot fix the phase factors of boundary wave functions
by this method, we can gain all the necessary informations of cylinder 
(annulus)
amplitudes and classify the Cardy states.

The most non-trivial point in the modular bootstrap for the boundary
$\cN=2$ Liouville theory is the following fact:
modular transformations of irreducible characters of the  $\cN=2$ SCA
with $\hat{c}(\equiv c/3)>1$ generically include the continuous spectrum of
$U(1)$-charges.  This feature is  not likely to fit to
any superstring vacuum on which the space-time supercharges act locally. 
Our key idea to overcome this difficulty 
is to consider the {\em sum over spectral flows\/} of 
the irreducible characters as the fundamental ingredients for the bootstrap.

This paper is organized as follows:
In section 2 we introduce the ``extended characters'' that are 
defined as suitable sums under spectral flow of irreducible characters and 
examine their modular transformation properties.
They are the basic conformal blocks in our modular bootstrap 
analysis. In section 3 we solve the modular bootstrap equations
and construct three types of candidate Cardy states, 
the ``class 1,2 and 3 states''. The class 1 branes are found to be   
natural generalizations of the ZZ-branes \cite{ZZ} in the bosonic
Liouville theory and localized at the strong coupling region 
$\phi\,\sim\,+\infty$. The class 2 and 3 branes are regarded as 
the analogues of the FZZT-branes \cite{FZZ,Teschner}, which extend 
along the Liouville direction.
In section 4 we apply our result to the superconformal systems  
describing the Calabi-Yau singularities mentioned above. 
The class 1 branes are identified with (the Liouville sector of)
BPS $D$-branes wrapped around vanishing cycles, and we show 
that the open string Witten indices reproduce the correct intersection
numbers among them. 
In section 5 we briefly discuss the non-BPS $D$-branes in the 
2d type 0 (and type II) string vacua composed only of the $\cN=2$
Liouville with $\hat{c}=5$. 
In section 6 we summarize the results of our analysis and 
make comments on some open issues.

~


\section{Modular Data for the $\cN=2$ Liouville Theory}

~

The $\cN=2$ Liouville theory \cite{KutS} is an $\cN=2$ $SCFT_2$ defined 
as the system of one chiral superfield $\Phi$ (one complex boson and 
one complex fermion) with the linear dilaton of the background charge 
$\cQ$ ($\hat{c}\equiv c/3=1+\cQ^2$ in our normalization). 
The real part of bosonic component of $\Phi$ is identified as  
the Liouville field $\phi$ and the imaginary part $Y$ is 
associated to the $U(1)$-current of $\cN=2$ SCA.
We have two types of the Liouville potentials preserving 
$\cN=2$ superconfomal symmetry; the chiral one $S_+$ which has  
the Liouville (imaginary) momentum $1/\cQ$ 
(we also have its anti-chiral counterpart $S_-$), 
and the non-chiral one $S_{\msc{nc}}$ which has the Liouville momentum $\cQ$;
\begin{eqnarray}
S_{\pm} &=& \int d^2z d^2\th^{\pm} \, e^{\frac{1}{\cQ}\Phi^{\pm}}~, 
\label{cosm chiral}\\
S_{\msc{nc}} &=& \int d^2z d^2\th^+d^2\th^-\,
    e^{\frac{\cQ}{2}(\Phi^+ + \Phi^-)} ~.
\label{cosm non-chiral} 
\end{eqnarray}

Since our interests are concentrated on cylinder amplitudes, 
the relevant conformal blocks should be expanded  by 
the irreducible characters of $\cN=2$ SCA. 
Hence we first  summarize the character formulas of 
unitary irreducible representations of $\cN=2$ SCA with $\hat{c}>1$ 
\cite{BFK,Dobrev}. We here only consider the NS sectors, but 
one can immediately derive the formulas for other spin structures 
by the half-integral spectral flows.
We denote the conformal weight and $U(1)$-charge of the highest weight state
as $h$, $Q$ and set $q\equiv e^{2\pi i\tau}$, $y\equiv e^{2\pi i z}$
  \footnote
    {We borrow the names of representations of $\cN=2$ SCA
     from \cite{ET,EOTY} where string compactifications on 
     Calabi-Yau manifolds have been studied.
     In the special case 
     $\hat{c}=5$ the string vacuum consists only of the $\cN=2$
      Liouville sector (together with the superconformal ghost sector), 
      and the ``graviton representation''  actually corresponds to a tachyon.};
\begin{description}
 \item[(1) massive representations] :
\begin{eqnarray}
\ch{(\sNS)}{}(h,Q;\tau,z) = q^{h-(\hat{c}-1)/8}\,y^Q\,
\frac{\theta_3(\tau,z)}{\eta(\tau)^3}~, ~~~(h>|Q|/2,~~0\leq |Q| 
< \hat{c}-1)~.
\label{massive character}
\end{eqnarray}

\item[(2) massless matter representations] : 
\begin{eqnarray}
\chm{(\sNS)}(Q;\tau,z)=q^{\frac{|Q|}{2}-(\hat{c}-1)/8}y^Q\,
\frac{1}{1+y^{\msc{sgn}(Q)}q^{1/2}}
\, \frac{\theta_3(\tau,z)}{\eta(\tau)^3}~.
\label{massless character 1}
\end{eqnarray}
They correspond to the (anti-)chiral primary state with $h=|Q|/2$, ~
($0<|Q|<\hat{c}$).

\item[(3) graviton representation] : 
\begin{eqnarray}
\chg{(\sNS)}(\tau,z)=q^{-(\hat{c}-1)/8}\,
\frac{1-q}{(1+y q^{1/2})(1+y^{-1}q^{1/2})} \,
\frac{\theta_3(\tau,z)}{\eta(\tau)^3}~.
\label{massless character 2}
\end{eqnarray}
They correspond to the vacuum $h=Q=0$, 
which is the unique state being both chiral and anti-chiral primary. 
\end{description}

More general unitary representations are generated by the integral 
spectral flows. Generally the spectral flow generators $U_{\eta}$
are defined with a real parameter $\eta$ by 
\begin{eqnarray}
&& U_{\eta}^{-1}L_m U_{\eta}
=L_m+\eta J_m+\frac{\hat{c}}{2}\eta^2\delta_{m,0}~,\nn
&& U_{\eta}^{-1}J_m U_{\eta} = J_m+ \hat{c}\eta \delta_{m,0}~,\nn
&& U_{\eta}^{-1}G^{\pm}_r U_{\eta}
= G^{\pm}_{r\pm \eta}~,
\label{spectral flow}
\end{eqnarray}
and we focus on the integer flow parameter $\eta=n\in \bz$ for the time being.
The corresponding  characters are given by 
\begin{eqnarray}
\ch{(\sNS)}{*}(*,n;\tau,z) \equiv 
q^{\frac{\hat{c}}{2}n^2}
y^{\hat{c}n}\, \ch{(\sNS)}{*}(*;\tau,z+n\tau)~, ~~~(n\in \bz)~,
\label{flowed character}
\end{eqnarray}
where $\ch{(\sNS)}{*}(*;\tau,z)$ is the abbreviated notation of
\eqn{massive character}-\eqn{massless character 2}.
Note that the flowed massive character
$\ch{(\sNS)}{}(h,Q,n;\tau,z)$ has the same form as \eqn{massive
character} with the shifts: 
\begin{eqnarray}
&& h\,\rightarrow\, h+Qn+\frac{\hat{c}-1}{2}n^2~, ~~~
Q\,\rightarrow\, Q+(\hat{c}-1)n~,
\end{eqnarray}
(note that $\theta_3(\tau,z+n\tau)=q^{-n^2/2}y^{-n}\theta_3(\tau,z)$). 
However, the massless characters
$\chm{(\sNS)}(Q,n;\tau,z)$, $\chg{(\sNS)}(n;\tau,z)$
differ from the original ones \eqn{massless character 1}, 
\eqn{massless character 2} in a non-trivial manner and possess
different structures of singular vectors.

The main purpose in this section is to clarify the modular transformation 
properties of the $\cN=2$ Liouville theory. 
We first point out that the $\cN=2$ Liouville theory is 
an $\cN=2$ SCFT of $\hat{c}>1$ {\em that possesses no
conserved quantities other than the conformal weight and
$U(1)$-charge\/}. (We emphasize that 
the Liouville field $\Phi$ is not free because of the Liouville potential
\eqn{cosm chiral}.).
This is analogous to the fact that the bosonic Liouville theory 
is a CFT with $c>1$ which has no continuous symmetry 
other than the Virasoro algebra.  
Therefore the relevant conformal blocks (for the computation of 
one-loop vacuum amplitudes) 
are expected to be the irreducible characters themselves.

However, this is not the whole story:
the modular transforms of the characters \eqn{massive
character}, \eqn{massless character 1}, \eqn{massless character 2}
and \eqn{flowed character} include continuous spectrum of 
$U(1)$-charges. This feature is not likely to be 
compatible with any superstring vacuum in which the space-time 
supercharges are well-defined as local operators. 
This difficulty originates from the irrationality of the relevant 
conformal system, and the prototype of their resolution
has been given in \cite{ET,EOTY,Odake}: {\em we should take 
the sums of irreducible characters over spectral flows and
use these ``extended" characters as the fundamental conformal blocks.} 
Extended characters possess integral $U(1)$-charges, 
good modular behaviors and are closed 
among themselves under modular transformations as in 
rational conformal field theories (but we need continuous spectrum of 
conformal weights). 
Although such extended characters are (infinitely) reducible in the sense of  
the $\cN=2$ SCA, they are irreducible from the point of view of the ``extended chiral
algebra'', which are defined by adding the spectral flow generators.
For example, adding the currents $J^{\pm}$ to the $\cN=2$ algebra at $\hat{c}=2$ gives 
the $\cN=4$ SCA of level 1 \cite{ET}.
We will summarize the modular properties of the extended
characters in the cases of $\hat{c}=2,3,4,5$ in Appendix C, which 
are the special examples of our analysis on the $\cN=2$ Liouville 
theory given below.


Our main interests in this paper are concentrated on
the $\cN=2$ Liouville theory in the cases of rational central 
charges $\hat{c}$.
For our later convenience we parameterize the background charge $\cQ$
as 
\begin{eqnarray}
\cQ=\sqrt{2K/N}~,~~~\hat{c}=1+\frac{2K}{N}~,
\end{eqnarray}
where $\dsp K\in
\bz_{>0}$, $N\in \bz_{>0}$. Different values of $N,K$ give different
theories even when $\hat{c}$ is the same, since it is found that
they have different spectra of $U(1)$-charge as we will clarify later. 

The models with irrational central charges are not likely to be  
relevant for the construction of superstring vacua, and are beyond the
scope of our study. However, one may also treat the irrational cases 
by taking the limit $N,K\,\rightarrow\,\infty$ with tuning  
the ratio $K/N$ properly, since the following analysis is clearly
applicable for arbitrary positive integers $N$, $K$. 
We will derive the modular transformation 
formulas relevant for this limit at the last of this section.

As stated above, we consider sums over spectral flows of
irreducible characters. 
However, there is still a subtlety due to the fractionality 
of central charge $\hat{c}$.
In fact, the full summation over integral spectral flows 
yields the theta functions at fractional level $K/N$ 
(see \eqn{extended character 1} in
Appendix C), which do not 
behave well under modular transformations. 
We thus take  the ``mod $N$'' {\em partial\/} sums over spectral flows,
which amounts to adding the spectral flow generators $U_{\pm N}$ (and 
its superpartners) to the chiral algebra. 
Essentially the same prescription has been used in \cite{ES1} 
in constructing the conformal blocks of toroidal partition function 
in singular Calabi-Yau manifolds. 

Thus we introduce the following extended characters 
which will be the basic ingredients for our modular 
bootstrap approach;
\begin{eqnarray}
\chi^{(\sNS)}(h_0,r,j_0;\tau,z)
&\equiv & 
\sum_{n\in r+N\bsz}\, q^{\frac{\hat{c}}{2}n^2}y^{\hat{c}n}\,
\ch{(\sNS)}{}\left(h_0,Q=\frac{j_0}{N};\tau,z+n\tau\right)~,  
\label{Liouville extended 0}\\
\chim^{(\sNS)}(r,s;\tau,z)
&\equiv& 
\sum_{n\in r+N\bsz}\, q^{\frac{\hat{c}}{2}n^2}y^{\hat{c}n}\,
\chm{(\sNS)}\left(Q=\frac{s}{N};\tau,z+n\tau\right)~,
\label{Liouville extended 0-2} \\
\chig^{(\sNS)}(r;\tau,z)
&\equiv & 
\sum_{n\in r+N\bsz}\, q^{\frac{\hat{c}}{2}n^2}y^{\hat{c}n}\,
\chg{(\sNS)}\left(\tau,z+n\tau\right)~,
\label{Liouville extended 0-3}
\end{eqnarray}
where the ranges of parameters $r$, $j_0$, $s$ are defined as
\begin{eqnarray}
&& r\in \bz_{N}~,~~~
0 \leq  j_0 \leq  2K-1~, ~~~
1 \le s \le N+2K-1~,
\label{range r s}
\end{eqnarray}
We shall assume $j_0$, $s$ are both integers to achieve the locality 
of spectral flow generators $U_{\pm N}$.
We later discuss the construction of superstring vacua by tensoring with 
rational conformal theories (especially, the $\cN=2$ minimal models)
and this assumption is necessary for the GSO projection to ensure
the integral total $U(1)$-charges.

In more physical viewpoints one may note that 
the spectral flow operators $U_{\pm N}$
generate particular winding modes along the $U(1)$-direction
and the assumption of their locality restricts the $U(1)$-charges
as in usual Kaluza-Klein modes. 
Different choices of the integer parameters $N$, $K$ 
yield different spectra of $U(1)$-charges
and correspond to different radii of the compact boson $Y$, 
even if $\hat{c}$ is equal, as we already mentioned. 
We will return this point in section 5 for a special example $\hat{c}=5$.

Let us present more explicit calculations;

\noindent
{\bf (1) extended massive characters}

For the massive characters \eqn{Liouville extended 0}
it is convenient to parameterize the characters by the conformal 
weight $h$ of the flowed vacuum at $n=r$, that is,
\begin{eqnarray}
h\equiv h_0+\frac{rj_0}{N}+\frac{Kr^2}{N}~,
\end{eqnarray}
and the summation \eqn{Liouville extended 0} boils down  to  
\begin{eqnarray}
 q^{h-\frac{(j_0+2Kr)^2+K^2}{4NK}}\, \Theta_{j_0+2Kr,NK}
\left(\tau,\frac{2z}{N}\right)
\, \frac{\theta_3(\tau,z)}{\eta(\tau)^3}~.
\end{eqnarray}
Note that the parameters $r$, $j_0$ appear 
only through the combination $j\equiv j_0+2Kr$. It is thus more convenient 
to define the extended massive characters by
\begin{eqnarray}
\chi^{(\sNS)}(h,j;\tau,z) =  q^{h-\frac{j^2+K^2}{4NK}}\,
\Th{j}{NK}\left(\tau,\frac{2z}{N}\right)\,\frac{\theta_3(\tau,z)}
{\eta(\tau)^3}~.
\label{Liouville massive}
\end{eqnarray}
The parameter $j$ runs over the range $j \in \bz_{2NK}$ and
the $U(1)$-charge of vacuum state is equal to $Q=j/N$.
By taking the spectral flows of the unitarity condition 
for $r=0$ given in \eqn{massive character}, $i.e.$ $h_0 \geq j_0/(2N)$,
one can readily find the unitarity condition for general cases \cite{BFK}
\begin{eqnarray}
h - \frac{j^2+K^2}{4NK} + \frac{(j_0-K)^2}{4NK} \geq 0~,
\label{unitarity massive}
\end{eqnarray}
where $j_0 $ is uniquely determined by the conditions
\begin{eqnarray}
0 \leq  j_0 \leq 2K-1 ~,~~~ j_0 \equiv j ~(\mod\, 2K)~.
\end{eqnarray}

~


\noindent
{\bf (2) extended massless characters}

For the massless matter representations \eqn{massless character 1}
the spectral flow sum \eqn{Liouville extended 0-2} becomes
\begin{eqnarray}
\chim^{(\sNS)}(r,s;\tau,z) 
&=& \sum_{n\in r+N\bsz}\,
 q^{\frac{\hat{c}}{2}n^2} y^{\hat{c}n} \, 
\chm{(\sNS)}\left(Q=\frac{s}{N};\tau,z+n\tau\right) \nn
&\equiv& \sum_{m\in \bsz}\, \frac{\left(yq^{N\left(m+\frac{2r+1}{2N}\right)}
\right)^{\frac{s-K}{N}}}{1+yq^{N\left(m+\frac{2r+1}{2N}\right)}}\,
y^{2K\left(m+\frac{2r+1}{2N}\right)} q^{NK\left(m+\frac{2r+1}{2N}\right)^2}
\,\frac{\theta_3(\tau,z)}{\eta(\tau)^3}~. 
\label{Liouville massless 1}
\end{eqnarray}
The corresponding vacuum state has the quantum numbers
\begin{eqnarray}
&& h=\frac{Kr^2+\left(r+\frac{1}{2}\right)s}{N} ~, 
~~ Q=\frac{s+2Kr}{N}~,~~ \mbox{for}~ 
0 \leq r <\frac{N}{2} ~,\nn
&& h=\frac{Kr^2-\left(r+\frac{1}{2}\right)(N-s)}{N} ~, 
~~ Q=\frac{s-N+2Kr}{N}~,~~ \mbox{for}~ 
-\frac{N}{2} \leq r \leq -1~.
\label{h Q massless matter}
\end{eqnarray}
Note that we have $\dsp h= \frac{Q}{2}=\frac{s}{2N}$ when $r=0$
(chiral primary) and also $\dsp h= -\frac{Q}{2}=\frac{N+2K-s}{2N}$
(anti-chiral primary) when $r=-1$.
We also remark ``charge conjugation'' relation
\begin{eqnarray}
\chim^{(\sNS)}(r,s;\tau,-z)= \chim^{(\sNS)}(-r-1,N+2K-s;\tau,z)~.
\end{eqnarray}


\noindent
{\bf (3) extended graviton characters}

The remaining extended massless characters \eqn{Liouville extended 0-3}
is calculated as
\begin{eqnarray}
\chig^{(\sNS)}(r;\tau,z)
&=& \sum_{n\in r+N\bsz}\, q^{\frac{\hat{c}}{2}n^2}y^{\hat{c}n}\,
\chg{(\sNS)}(\tau,z+n\tau) \nn
&\equiv &
q^{-\frac{K}{4N}}\,\sum_{m\in\bsz}\, 
q^{NK\left(m+\frac{r}{N}\right)^2+N\left(m+\frac{2r-1}{2N}\right)} 
y^{2K\left(m+\frac{r}{N}\right)+1}\, \nn
&& \hspace{1cm} \times
\frac{1-q}{\left(1+yq^{N\left(m+\frac{2r+1}{2N}\right)}\right)
\left(1+yq^{N\left(m+\frac{2r-1}{2N}\right)}\right)}
\, \frac{\theta_3(\tau,z)}{\eta(\tau)^3}~. 
\label{Liouville massless 2}
\end{eqnarray}
We can show the following character identity
\begin{eqnarray}
&&\chi^{(\sNS)}\left(h=0,j=0;\tau,z\right)
=\chig^{(\sNS)}(r=0;\tau,z) \nn
&& \hspace{1cm} 
+\chim^{(\sNS)}(r=0,s=N;\tau,z)
+\chim^{(\sNS)}(r=-1,s=2K;\tau,z)~.
\label{Liouville character relation 2}
\end{eqnarray}
Similar character identities have been already 
known in \cite{ET,Odake,HS} in the special examples $\hat{c}=2,3,4$.
More generally, we obtain by taking the spectral flow of
\eqn{Liouville character relation 2}
\begin{eqnarray}
&&\chi^{(\sNS)}\left(h=\frac{Kr^2}{N},j=2Kr;\tau,z\right)
=\chig^{(\sNS)}(r;\tau,z) \nn
&& \hspace{2cm} 
+\chim^{(\sNS)}(r,s=N;\tau,z)
+\chim^{(\sNS)}(r-1,s=2K;\tau,z)~,
\label{Liouville character relation}
\end{eqnarray}
which is quite useful and will be often used in the later analysis.

~

Now, we are in a position to 
present the modular transformation formulas which are crucial 
to our modular bootstrap program. 
We shall only consider the S-transformations of the NS characters,
and use the following abbreviated notations
\begin{eqnarray}
&& \chi(p,j;\tau,z) \equiv \chi^{(\sNS)}\left(
h= \frac{p^2}{2}+\frac{j^2+K^2}{4NK}, j;\tau,z
\right) \equiv q^{p^2/2}\Theta_{j,NK}\left(\tau,\frac{2z}{N}\right)
\,\frac{\theta_3(\tau,z)}{\eta(\tau)^3}
~, 
\label{abbreviated 1}
\\
&& \chim^{}(r,s;\tau,z) \equiv \chim^{(\sNS)}(r,s;\tau,z)~,~~~
\chig^{}(r;\tau,z) \equiv \chig^{(\sNS)}(r;\tau,z)~.
\label{abbreviated 2}
\end{eqnarray}
Note that the real ``momentum'' $p$ here means  
that no tachyons appear in the spectrum (as is obvious from the last 
expression in \eqn{abbreviated 1}). 
On the other hand, we note that the unitarity condition 
\eqn{unitarity massive} allows a range of 
pure imaginary values of $p$;
\begin{eqnarray}
\frac{p^2}{2} \geq - \frac{(j_0-K)^2}{4NK}~,~~
 j_0 \equiv j ~,~  (\mod\, 2K)~, ~~  0\leq j_0 \leq 2K-1~.
\label{unitarity bound class 2}
\end{eqnarray} 

~

\noindent
{\bf (1) extended massive characters}

The S-transformation of the extended massive characters 
\eqn{Liouville massive} is easy. It is reduced to the Fourier transformation 
of Gaussian integral and the familiar modular properties of 
theta function with the level $NK$; 
\begin{eqnarray}
&& \chi\left(p, j ; -\frac{1}{\tau}, \frac{z}{\tau}\right)  
= e^{i\pi\frac{\hat{c} z^2}{\tau}} \sqrt{\frac{2}{NK}}\,
\sum_{j' \in \bsz_{2NK}}\, e^{-2\pi i \frac{j j'}{2NK}} 
\, \int_{0}^{\infty}dp'\,
 \cos(2\pi p p') \,
\chi\left(p', j' ; 
\tau,z \right)~. \nn
&&
\label{Liouville massive S 1}
\end{eqnarray}

~

\noindent
{\bf (2) extended massless characters}

The S-transformation of the extended massless character \eqn{Liouville
massless 1} is quite non-trivial. \\

\noindent
{\bf Under the S-transformation
the massless representations
become the sum of massive and massless representations}  
\cite{ET,Odake,Miki}.\\

Fortunately a useful  formula for the relevant calculations 
has been provided  in \cite{Miki}. We exhibit  it in Appendix B.
We first note the relation;
\begin{eqnarray}
\chim(r,s;\tau,z)= \frac{1}{N}\sum_{j=0}^{N-1}\,
e^{-2\pi i \frac{(2r+1)j}{2N}}\, I\left(2K/N,j/N,(s-K)/N;\tau,z\right)
\, \frac{\theta_3(\tau,z)}{\eta(\tau)^3}~,
\end{eqnarray}
where $I(k,a,b;\tau,z)$ is defined in \eqn{Ikab}. 
Using the formula \eqn{Miki formula}, 
we can derive the desired modular transformation formula\footnote
 {Since the massless characters can be reexpressed as alternating
  infinite sums of the massive characters, one may wonder how 
  the modular transforms  \eqn{Liouville massless S 1},
  \eqn{Liouville massless S 2} contain the massless pieces. 
  However, naive infinite sums of the modular coefficients for 
  the massive representations lead to divergences. Mathematically 
  well-defined treatments for the relevant calculus are found  in 
  \cite{Miki}, and the ``contour deformation technique'' used there
  yields extra pole contributions which generate the massless characters. 
  We also note that the last line in  \eqn{Liouville massless S 1},
  which may appear somewhat peculiar, originates from poles 
  on the real axis of $p'$ (and the $p'$-integral in the first line
  should be interpreted as the principal value).  
  Similar arguments are also found in \cite{Teschner} for the
  calculations of cylinder amplitudes.}
;
\begin{eqnarray}
&&\chim\left(r,s;-\frac{1}{\tau}, \frac{z}{\tau}\right)
=e^{i\pi \frac{\hat{c} z^2}{\tau}} \, \left\lb
\frac{1}{\sqrt{2NK}} \,\sum_{j'\in \bsz_{2NK}}\,
e^{-2\pi i \frac{(s+2Kr)j'}{2NK}}\,
\int_{0}^{\infty} dp' \, \right. \nn
&& \hspace{1cm}\times
\frac{\cosh\left(2\pi \frac{N+K-s}{N}\frac{p'}{\cQ}\right)
+e^{i\frac{\pi j'}{K}} \cosh\left(2\pi \frac{s-K}{N}\frac{p'}{\cQ}\right) }
{2\left|\cosh \, \pi \left(\frac{p'}{\cQ}+i\frac{j'}{2K}\right)\right|^2}
\, \chi\left(p', j';\tau,z\right)
\nn
&& \hspace{1cm}
+\frac{i}{N}
\sum_{r'\in\bsz_{N}}\,\sum_{s'=K+1}^{N+K-1}\,
e^{-2\pi i \frac{(s+2Kr)(s'+2Kr')-(s-K)(s'-K)}{2NK}}\, \chim(r',s';\tau,z)\nn
&& \hspace{1cm} \left.
+\frac{i}{2N}\,\sum_{r'\in\bsz_{N}} 
\,e^{-2\pi i \frac{(s+2Kr)(2r'+1)}{2N}}\,
\left\{\chim(r',K;\tau,z)-\chim(r',N+K;\tau,z)\right\}
\right\rb ~.
\label{Liouville massless S 1}
\end{eqnarray}

\noindent
{\bf (3) extended graviton characters}

For the graviton representations \eqn{Liouville massless 2},
we also obtain 
\begin{eqnarray}
&&\chig^{}\left(r;-\frac{1}{\tau}, \frac{z}{\tau}\right)
=e^{i\pi \frac{\hat{c} z^2}{\tau}} \, \left\lb
\frac{1}{\sqrt{2NK}} \,\sum_{j'\in \bsz_{2NK}}\,
e^{-2\pi i \frac{rj'}{N}}\, 
\int_{0}^{\infty} dp' \, 
\frac{\sinh\left(\pi \cQ p'\right)\sinh\left(2\pi \frac{p'}{\cQ}\right)}
{\left|\cosh \, \pi \left(\frac{p'}{\cQ}+i\frac{j'}{2K}\right)\right|^2}\, 
\chi\left(p', j';\tau,z\right) \right.
\nn
&& \hspace{1cm}
\left.
+\frac{2}{N}
\sum_{r'\in \bsz_{N}}\,\sum_{s'=K+1}^{N+K-1}\,
\sin\left(\frac{\pi (s'-K)}{N}\right) e^{-2\pi i \frac{r(s'+2Kr')}{N}}
\, \chim^{}(r',s';\tau,z)\right\rb ~,
\label{Liouville massless S 2}
\end{eqnarray}
with the help of the character relation  \eqn{Liouville character relation}.
It is important to note that the parameter $s'$ appearing in 
the R.H.S's of \eqn{Liouville massless S 1} and 
\eqn{Liouville massless S 2} runs only over the {\em partial\/} 
range $K\leq s' \leq N+K$, 
instead of the full range of the unitary (matter) representations 
$1\leq s'\leq  N+2K-1$. Moreover, graviton representations
\eqn{Liouville massless 2} do not appear at all in the R.H.S's of 
\eqn{Liouville massless S 1} and 
\eqn{Liouville massless S 2}.
These features are in a sharp contrast with 
rational conformal field theories, and have been already found in 
the $\hat{c}=2$ case \cite{ET} (see Appendix C). 
We also point out  that all the massless matter representations in this range 
$K\leq s' \leq N+K$ only include 
states above the mass gap $\cQ^2/8 \equiv K/(4N)$,
as is easily checked from the formulas \eqn{h Q massless matter}.

To close this section let us consider the limit $N\,\rightarrow\,
\infty$, $K\,\rightarrow\,\infty$ with keeping the value $\cQ^2=2K/N$ fixed.
Under this limit the sums over spectral flows \eqn{Liouville extended 0},
\eqn{Liouville extended 0-2}, \eqn{Liouville extended 0-3}
should be replaced with the original irreducible characters 
\eqn{flowed character}
with the identifications $r=n (\in \bz)$, ~ $j/N= \om$~ 
($-\infty < \om < +\infty$), ~ $s/N=\la$ ~ ($0<\la<\hat{c}(\equiv 1+\cQ^2) $);
\begin{eqnarray}
&& \chi^{(\sNS)}(h,j;\tau,z)~ \longrightarrow~
\ch{(\sNS)}{}(h,\om;\tau,z) (\equiv \ch{(\sNS)}{}(h_0,\om_0,n;\tau,z))
\nn
&& (\mbox{where} ~ \om=\om_0+\cQ^2 n~, ~~0\leq \om_0 < \cQ^2~,~~~
  h= h_0+ \om_0 n + \frac{\cQ^2}{2} n^2 )\nn
&& \chim^{(\sNS)}(r,s;\tau,z) ~ \longrightarrow~
\chm{(\sNS)}(\la,n;\tau,z)~,~~~
\chig^{(\sNS)}(r;\tau,z) ~ \longrightarrow~
\chg{(\sNS)}(n;\tau,z)~.
\end{eqnarray}
The modular transformation formulas are reduced to
\begin{eqnarray}
&& \ch{(\sNS)}{}\left(\frac{p^2}{2}+\frac{\om^2}{2\cQ^2}+\frac{\cQ^2}{8}, \om
 ;  -\frac{1}{\tau}, \frac{z}{\tau}\right) 
= e^{i\pi\frac{\hat{c} z^2}{\tau}} \frac{2}{\cQ}\,
\int_{-\infty}^{\infty} d\om'\, \int_{0}^{\infty}dp'\, \nn
&& \hspace{4cm} \times
e^{-2\pi i \frac{\om \om'}{\cQ^2}} \, 
 \cos(2\pi p p') \,
\ch{(\sNS)}{}\left(\frac{p^{'2}}{2}+\frac{\om^{'2}}{2\cQ^2}+\frac{\cQ^2}{8},
\om' ; \tau,z \right)~,
\label{Liouville massive S cont} \\
\vspace{5mm}
&&\chm{(\sNS)}\left(\la,n;-\frac{1}{\tau}, \frac{z}{\tau}\right)
=e^{i\pi \frac{\hat{c} z^2}{\tau}} \, \left\lb
\frac{1}{\cQ} \, \int_{-\infty}^{\infty}d\om'\, \int_{0}^{\infty} dp' \, 
e^{-2\pi i \frac{(\la+\cQ^2 n)\om'}{\cQ^2}}\,
 \right. \nn
&& \hspace{5mm}\times
\frac{\cosh\left(2\pi \left(1+\frac{\cQ^2}{2}-\la\right)\frac{p'}{\cQ}\right)
+e^{2\pi i\frac{\om'}{\cQ^2}} 
\cosh\left(2\pi \left(\la-\frac{\cQ^2}{2}\right)\frac{p'}{\cQ}\right) }
{2\left|\cosh \, \pi \left(\frac{p'}{\cQ}+i\frac{\om'}{\cQ^2}\right)\right|^2}
\, \ch{(\sNS)}{}\left(\frac{p^{'2}}{2}
+\frac{\om^{'2}}{2\cQ^2}+\frac{\cQ^2}{8}, \om';\tau,z\right)
\nn
&& \hspace{5mm} \left.
+i\sum_{n'\in \bsz}\,\int_{\frac{\cQ^2}{2}}^{1+\frac{\cQ^2}{2}}d\la' \,
e^{-2\pi i \frac{(\la+\cQ^2n)(\la'+\cQ^2 n')-
   \left(\la-\frac{\cQ^2}{2}\right)\left(\la'-\frac{\cQ^2}{2}\right)}{\cQ^2}}
\, \chm{(\sNS)}(\la',n';\tau,z) \right\rb ~,
\label{Liouville massless S 1 cont} \\
\vspace{5mm}
&&\chg{(\sNS)}\left(n;-\frac{1}{\tau}, \frac{z}{\tau}\right)
=e^{i\pi \frac{\hat{c} z^2}{\tau}} \, \left\lb
\frac{1}{\cQ} \,\int_{-\infty}^{\infty} d\om' \,\int_{0}^{\infty} dp'\,
e^{-2\pi i n\om'}\, \right.  \nn
&& \hspace{5mm} \times
\frac{\sinh\left(\pi \cQ p'\right)\sinh\left(2\pi \frac{p'}{\cQ}\right)}
{\left|\cosh \, \pi \left(\frac{p'}{\cQ}+i\frac{\om'}{\cQ^2}\right)\right|^2}
\, \ch{(\sNS)}{}\left(\frac{p^{'2}}{2}
+\frac{\om^{'2}}{2\cQ^2}+\frac{\cQ^2}{8}, \om';\tau,z\right) 
\nn
&& \hspace{5mm}
\left.
+2\sum_{n'\in \bsz}\,\int_{\frac{\cQ^2}{2}}^{1+\frac{\cQ^2}{2}} d\la' \,
\sin\left(\pi \left(\la'-\frac{\cQ^2}{2}\right)\right) 
e^{-2\pi i n \left(\la'+\cQ^2 n'\right)}
\, \chm{(\sNS)}(\la',n';\tau,z)\right\rb ~.
\label{Liouville massless S 2 cont}
\end{eqnarray}
These formulas are
relevant for the $D$-branes in the type 0 string vacua, 
where we need not impose the integrality of $U(1)$-charge, 
as we will see in our later discussions.

~

\section{Modular Bootstrap Approach to $\cN=2$ Boundary \\ Liouville Theory}

\subsection{Modular Bootstrap : Two Examples}

~

To begin with, we have to clarify what the ``modular bootstrap'' means. 
We present two helpful examples for this purpose.
In general (super)conformal field theories are not necessarily rational and
there exists a continuous spectrum of primary fields: let us label continuous representations 
by a parameter $p$ (``momentum'') and label the discrete 
representations by a parameter $I$. 
We denote the corresponding characters of the chiral algebra
as $\chi_p(\tau)$ and $\chi_I(\tau)$, respectively. The Ishibashi states
\cite{Ishibashi} are defined by imposing suitable boundary conditions on 
the chiral algebra 
including the Virasoro algebra;
\begin{eqnarray}
 (L_n-\tL_{-n})\dket{p}=0~,~~~ (L_n-\tL_{-n})\dket{I}=0~,
\end{eqnarray} 
and determined uniquely (up to a phase factor)  
by the orthonormality conditions as follows;
\begin{eqnarray}
&& \dbra{p}e^{-\pi T H^{(c)}}\dket{p'}= \delta(p-p')\chi_p(iT)~,~~~
\dbra{I}e^{-\pi T H^{(c)}}\dket{I'}= \delta_{I,I'}\chi_I(iT)~, \nn
&& \dbra{p}e^{-\pi T H^{(c)}}\dket{I}=0~,
\end{eqnarray}
where $\dsp H^{(c)}\equiv L_0+\tL_0-\frac{c}{12}$ 
is the closed string Hamiltonian and 
$\tilde{q}\equiv e^{-2\pi T}$ is the closed string modulus for 
the cylinder amplitude.

Consistent $D$-branes are described by 
the Cardy states  \cite{Cardy} of the form 
\begin{eqnarray}
\ket{B;\xi} 
= \int dp\, \Psi_{\xi}(p)\dket{p} 
+ \sum_I\, C_{\xi}(I)\dket{I} ~,
\end{eqnarray}
which should satisfy the following condition 
exhibiting the open-closed string duality
\begin{eqnarray}
\bra{B;\xi_1} e^{-\pi T H^{(c)}} \ket{B;\xi_2}
=\int dp\, \rho(p|\xi_1,\xi_2) \chi_p(it) + \sum_I N(I|\xi_1,\xi_2) 
\chi_I(it)~.
\label{Cardy condition}
\end{eqnarray}
In this expression $t\equiv 1/T$ means the open string modulus.
The ``spectral density''  $\rho(p|\xi_1,\xi_2)$ is a positive generalized 
function  (or distribution) 
and $N(I|\xi_1,\xi_2)$ are positive integers. 
Note that the positivity of $\rho(p|\xi_1,\xi_2)$ and 
$N(I|\xi_1,\xi_2)$ yields non-trivial constraints in general.

At first glance, solving the Cardy condition \eqn{Cardy condition}
appears to be a highly non-trivial problem. We have much more 
constraints than the number of unknowns. The best we can do is 
to solve this condition introducing a suitable Ansatz. 
To proceed further we discuss two examples:

~

\noindent
{\bf 1.}  The simplest example is the $SU(2)_k$ WZW model. 
We only have a discrete spectrum corresponding to the integrable
representations of the $SU(2)_k$ current algebra labelled by $\ell$ $(\ell=0,1,\ldots, k)$
and the modular data are given as
\begin{eqnarray}
&& \chi^{(k)}_{L} \left(-1/\tau, z/\tau\right) = 
   e^{i\pi \frac{k}{k+2}\frac{z^2}{\tau}}\,
\sum_{\ell=0}^{k}\, S^{(\msc{d})}(\ell|L)  
    \chi^{(k)}_{\ell}(\tau,z)~,~~~ \nn
&& S^{(\msc{d})}(\ell|L) = \sqrt{\frac{2}{k+2}} 
\sin\left(\frac{\pi(\ell+1)(L+1)}{k+2}\right)~,
\label{modular data SU(2)} 
\end{eqnarray} 
where the $SU(2)_k$ character $\chi^{(k)}_{\ell}(\tau,z)$ is defined in 
\eqn{SU(2) character}. 
Ishibashi states $\dket{\ell}$ are defined by
\begin{eqnarray}
(J^a_n+\tJ^a_{-n})\dket{\ell}=0~,~~~ \dbra{\ell} e^{-\pi T H^{(c)}} 
e^{i\pi z(J^3_0-\tJ^3_0)} \dket{\ell'} = \delta_{\ell,\ell'} 
\chi_{\ell}(iT,z)~,
\label{Ishibashi SU(2)}
\end{eqnarray}
and the Cardy states 
$
\dsp \ket{B;L}=\sum_{\ell}C_L(\ell)\dket{\ell}
$
are in a one-to-one correspondence with the integrable representations
($0\leq \ell, \, L \leq k$) \cite{Cardy}.  
We note that the Cardy states 
$\dsp \ket{B;L}$
are completely characterized up to phase factors by the following 
equations, which we would like to call 
as the ``modular bootstrap equations'',  
\begin{eqnarray}
e^{\pi \frac{k}{k+2}\frac{z^2}{T}}
\bra{B;0} e^{-\pi T H^{(c)}}e^{i\pi z(J^3_0-\tJ^3_0)} 
\ket{B;L} &=& \chi_{L}(it,z')~, ~~~({}^{\forall} L,~
t\equiv 1/T, ~ z'\equiv -it z)~.
\label{BO eq 1 SU(2)} 
\end{eqnarray}

One can easily solve the above equations with an Ansatz 
$C_{L}(\ell)=g(\ell)S^{(\msc{d})}(\ell|L)$,
where $g(\ell)$ are unknown coefficients independent of $L$.
In fact, we readily obtain from \eqn{BO eq 1 SU(2)},
\begin{eqnarray}
g(\ell) = 
\frac{1}{\sqrt{S^{(\msc{d})}(\ell|0)}}~,~~~
C_{L}(\ell) = \frac{S^{(\msc{d})}(\ell|L)}
{\sqrt{S^{(\msc{d})}(\ell|0)}}~,
\end{eqnarray} 
with a possible phase factor. 
As is well-known,
these solutions satisfy the Cardy condition \eqn{Cardy
condition} thanks to the Verlinde formula.
Note that the special Cardy state $\ket{B;0}$ is characterized 
by the equation 
\begin{eqnarray}
e^{\pi \frac{k}{k+2}\frac{z^2}{T}}
\bra{B;0} e^{-\pi T H^{(c)}}e^{i\pi z(J^3_0-\tJ^3_0)} 
 \ket{B;0} &=& \chi_{0}(it,z')~,
~~~(t\equiv 1/T, ~ z'\equiv -it z)~.
\label{BO eq 2 SU(2)} 
\end{eqnarray}
Since the primary field with $\ell=0$ is the identity,
this condition implies that the $D$-brane described by $\ket{B;0}$
only includes the identity as the boundary operator.
We obtain 
\begin{eqnarray}
C_{0}(\ell) = \sqrt{S^{(\msc{d})}(\ell|0)}~.
\end{eqnarray}

Similar constructions can be found in general rational 
conformal field theories,
and the modular bootstrap approach reproduces the correct 
Cardy states.

~

\noindent
{\bf 2.} 
A more non-trivial example is the boundary Liouville theory.
It is known that there exist Cardy states corresponding to both the non-degenerate
(``FZZT-branes'' \cite{FZZ,Teschner})
and degenerate representations (``ZZ-brane'' \cite{ZZ}). 
However, the 
Ishibashi states are only given by the non-degenerate representations
above the mass gap.  
The general Cardy states are written in the form 
\begin{eqnarray}
\ket{B;\xi} = \int_0^{\infty}dp \, \Psi_{\xi}(p)\dket{p}~,
\end{eqnarray}
where $\xi$ labels representations of Virasoro algebra and 
$\dket{p}$ denotes the Virasoro Ishibashi state with the conformal 
weight $\dsp h=p^2+\frac{\cQ^2}{4}$ ($p>0$). (We here denote the (bosonic) Liouville 
background charge as $\dsp \cQ\equiv b+\frac{1}{b}$, $c=1+6\cQ^2$, and 
take the convention $\alpha'=1$.)

The modular data for characters are written in the form as 
\begin{eqnarray}
\chi_{\xi} \left(-1/\tau\right) = \int_0^{\infty} dp'\, 
S^{(\msc{c})}(p'|\xi) \chi_{p'}(\tau)~,
\label{modular data Liouville} 
\end{eqnarray} 
where the coefficients $S^{(\msc{c})}(p'|\xi)$ are specified 
below. 
Note that the R.H.S of the above formula contains only the continuous 
series above the mass gap $h\ge \cQ^2/4$.

As in the previous example, Cardy states are 
characterized by the equations 
\begin{eqnarray}
\bra{B;O} e^{-\pi T H^{(c)}}
\ket{B;\xi} &=& \chi_{\xi}(it)~, 
\label{BO eq 1 Liouville} \\
\bra{B;O} e^{-\pi T H^{(c)}}
 \ket{B;O} &=& \chi_{h=0}(it)~,
\label{BO eq 2 Liouville} 
\end{eqnarray}
In \eqn{BO eq 2 Liouville}, 
$\chi_{h=0}(it)$ denotes the character of the $h=0$ state (identity
operator);
\begin{eqnarray}
&&\chi_{h=0}(\tau) \equiv
\chi_{1,1}(\tau) 
= \frac{1}{\eta(\tau)} \, 
\left(q^{-\frac{1}{4}\left(\frac{1}{b}+b\right)^2} -
q^{-\frac{1}{4}\left(\frac{1}{b}-b\right)^2}  \right)~.
\end{eqnarray}
The condition \eqn{BO eq 2 Liouville} again means that the 
$D$-brane associated to $\ket{B;O}$ only includes the boundary identity 
operator. It is identified as 
the ``(1,1)-brane'' (a special type of ZZ-brane) 
defined by the wave function \cite{ZZ}
\begin{eqnarray}
\Psi_O(p) =- 2^{5/4}\cdot 2\pi i p \hmu^{ip/b} \frac{1}{\Gamma(1+i2bp)
\Gamma\left(1+i\frac{2p}{b}\right)}~,
\label{wave function 11}
\end{eqnarray}
where $\hmu\equiv \pi \mu \gamma(b^2)$ ($\gamma(x)\equiv 
\Gamma(x)/\Gamma(1-x)$) is the ``renormalized" cosmological constant.
It indeed satisfies the equation \eqn{BO eq 2 Liouville}.
General Cardy states $\ket{B;\xi}$ are determined by the equation
\eqn{BO eq 1 Liouville} as 
\begin{eqnarray}
&& f(p) = 1/ \Psi_O(p)^{*} \equiv 2^{-5/4} \frac{1}{2\pi i p} 
\hmu^{ip/b} \Gamma(1-i2bp)\Gamma\left(1-i\frac{2p}{b}\right)~, \nn
&& \Psi_{\xi}(p) = f(p) S^{(\msc{c})}(p|\xi)~.
\label{wave function Liouville}
\end{eqnarray}

As for the FZZT-brane we obtain for the modular coefficients 
$S^{(\msc{c})}(p|\xi)$,
\begin{eqnarray}
S^{(\msc{c})}(p|s) = 2\sqrt{2}\cos(2\pi sp)~.
\end{eqnarray}
In other words $|B;s\rangle$ corresponds to the non-degenerate
representation with $h=s^2/4+{\cal Q}^2/4$. 
Based on the bootstrap of disc amplitudes and the perturbative analysis
of the Liouville theory it is possible to show that the parameter $s$ is related to
the boundary cosmological constant $\mu_B$ 
as $\dsp \cosh^2(\pi b s) =\frac{\mu_B^2}{\mu}\sin(\pi b^2)$ \cite{FZZ}.

We also obtain 
\begin{eqnarray}
S^{(\msc{c})} (p|m,n)
= 4\sqrt{2}\sinh\left(2\pi \frac{mp}{b}\right)\sinh(2\pi b n p)~,~~~
(m,n \in \bz_{\geq 0})~,
\end{eqnarray}
for the general ZZ-brane corresponding 
to the $(m,n)$-degenerate representation \\
($\dsp h=-\frac{1}{4}\left(\frac{m}{b}+nb\right)^2+\frac{Q^2}{4}$).
We note that the relation
\begin{eqnarray}
 \Psi_{m,n}(p)= \Psi_{s=i\left(\frac{m}{b}+nb \right)}(p)
- \Psi_{s=i\left(\frac{m}{b}-nb \right)}(p)~,
\end{eqnarray}
follows from the character formula 
\begin{eqnarray}
\chi_{m,n}(\tau) = \frac{1}{\eta(\tau)} \, 
\left(q^{-\frac{1}{4}\left(\frac{m}{b}+nb\right)^2} -
q^{-\frac{1}{4}\left(\frac{m}{b}-nb\right)^2}  \right)
 \equiv \chi_{p= \frac{i}{2}\left(\frac{m}{b}+nb \right)} (\tau)
 - \chi_{p= \frac{i}{2}\left(\frac{m}{b}-nb \right)} (\tau)~.
\end{eqnarray}

It is found that the Cardy condition \eqn{Cardy condition}
is in fact satisfied with the positive spectral densities (under a 
suitable regularization) at least among the FZZT-branes
above the mass gap $s\in \br_+$ and the general ZZ-brane.
However, the $(m,n)$-type ZZ-branes ($(m,n)\neq (1,1)$)
generally produce non-unitary representations in the open string 
channel,  and would not be regarded as  proper boundary states in  
string theory.

We emphasize that the solutions \eqn{wave function 11},
\eqn{wave function Liouville} can be reproduced 
except for the phase factors from the simple equations
\eqn{BO eq 1 Liouville}, \eqn{BO eq 2 Liouville} with 
a natural Ansatz
$\Psi_{\xi}(p)=f(p)S^{(\msc{c})}(p|\xi)$.
Consequently, as long as we work with cylinder amplitudes, 
the modular bootstrap approach provides all the necessary
information and we can classify the Cardy states.

~


\subsection{Modular Bootstrap in Boundary $\cN=2$ Liouville Theory and 
Classification of Cardy States}

~

Now, let us study the Cardy states in $\cN=2$ Liouville theory based 
on the modular bootstrap. 
It is well-known that the $\cN=2$ superconformal symmetry allows 
the following two types of boundary conditions \cite{OOY};
\begin{eqnarray}
&&  \mbox{\bf A-type}~~~ :~
(J_n-\tJ_{-n})\ket{B}=0 ~,~~~(G^{\pm}_r-i\tG^{\mp}_{-r})\ket{B}=0~,
\label{A-type} \\
&& \mbox{\bf B-type}~~~ :~
 (J_n+\tJ_{-n})\ket{B}=0 ~,~~~(G^{\pm}_r-i\tG^{\pm}_{-r})\ket{B}=0~,
\label{B-type}
\end{eqnarray}
Both of them are compatible with the $\cN=1$ superconformal symmetry
\begin{eqnarray}
(L_n-\tL_{-n})\ket{B}=0~,~~~ (G_r-i\tG_{-r})\ket{B}=0~,
\end{eqnarray}
where $G=G^++G^-$ is the $\cN=1$ supercurrent.
The following analysis is completely parallel for both A and 
B-type branes and we shall only consider the A-type from now on.

Encouraged by the successes of the previous examples, 
let us now discuss our modular bootstrap equations for ${\cal N}=2$ Liouville theory.
As addressed  before, we shall take the extended characters 
\eqn{Liouville massive},
\eqn{Liouville massless 1} and 
\eqn{Liouville massless 2} as the fundamental conformal blocks
defining the modular bootstrap equations.
This choice is quite natural in realizing the BPS $D$-branes in supersymmetric 
string vacua, as we already explained. 
They have a modular transformation \eqn{Liouville massive S 1},
\eqn{Liouville massless S 1} and  \eqn{Liouville massless S 2},
of the generic form
\begin{eqnarray}
\chi_{\xi} \left(-1/\tau, z/\tau\right) = e^{i\pi \frac{\hat{c}z^2}{\tau}}\,
\left\lb
\int_0^{\infty} dp'\, \sum_{j'}\, 
S^{(\msc{c})}(p', j'|\xi) \chi(p',j';\tau,z)
  +\sum_{r',s'}\, S^{(\msc{d})}(r',s'|\xi) 
    \chim^{~}(r',s';\tau,z)\right\rb~.~\nn
&&
\label{modular data 0} 
\end{eqnarray} 
In contrast to the previous examples, the modular transform includes
{\em both\/} continuous and discrete representations in general.

We now propose an Ansatz that the (A-type) Ishibashi states are spanned 
by the massive and the following massless matter 
representations, which form the maximal family closed under 
modular transformations;
\begin{eqnarray}
&&\dket{p,j}~~(p>0~,~~ j\in \bz_{2NK})~,~~~\nonumber \\
&&\dketm{r,s}~~(r\in {\bf Z}_N~,~~ s=K+1, \ldots, K+N-1)~,
\label{Ishibashi N=2 Liouville}
\end{eqnarray}
which satisfy the orthogonality conditions
\begin{eqnarray}
&& \dbra{p,j} e^{-\pi T H^{(c)}} e^{i\pi z(J_0+\tJ_0)}\dket{p',j'} 
= \delta(p-p')\delta^{(2NK)}_{j,j'}\, \chi(p,j;iT, z)~, \nn
&& \dbram{r,s} e^{-\pi T H^{(c)}} e^{i\pi z(J_0+\tJ_0)}\dketm{r',s'} 
= \delta^{(N)}_{r,r'}\delta_{s,s'}\, \chim^{}(r,s;iT, z)~, \nn
&&  \dbra{p,j}e^{-\pi T H^{(c)}} e^{i\pi z(J_0+\tJ_0)}\dketm{r,s}=0~, 
\label{Ishibashi N=2 Liouville OC}
\end{eqnarray}
Here the symbol $\delta^{(M)}_{m,m'}$ means the Kronecker delta mod. $M$.

Note that among the Ishibashi states
\begin{enumerate}
\item we do not include any of the graviton representations,
\item we include massless matter representations only 
in the range $K+1\le s\le K+N-1$.
\end{enumerate}
It is natural that the graviton representations do not appear 
in the closed string channel since gravity 
is decoupled in Liouville theory where 
all physical states in the closed string sector
exist above the mass gap. The graviton representations, however, do 
appear in the open string channel and the corresponding Cardy states 
describe basic $D$-branes 
of ${\cal N}=2$ Liouville theory as we shall see below.

Now we postulate 
the modular bootstrap equation for ${\cal N}=2$ Liouville theory
\footnote
    {If we take the B-type boundary conditions, 
    we instead make an insertion of $e^{i\pi z(J_0-\tJ_0)}$.}
\begin{eqnarray}
e^{\pi \frac{\hat{c}z^2}{T}}
\bra{B;O} e^{-\pi T H^{(c)}} e^{i\pi z(J_0+\tJ_0)}
\ket{B;\xi} &=& \chi_{\xi}(it,z')~, 
\label{BO eq 1} \\
e^{\pi \frac{\hat{c}z^2}{T}}
\bra{B;O} e^{-\pi T H^{(c)}} e^{i\pi z(J_0+\tJ_0)}\ket{B;O} 
&=& \chig^{~}(r=0;it,z')~,~~~
(T\equiv 1/t,~z'\equiv -it z)
\label{BO eq 2}
\end{eqnarray}
where $\chig^{~}(r=0;it,z')$ is the character
for the graviton  representation $h=Q=0$. 
General Cardy states have the form
\begin{eqnarray}
&& \ket{B;\xi} = \int_0^{\infty}dp\, \sum_{j \in \bsz_{2NK}}\,
\Psi_{\xi}(p,j)\dket{p,j}+ 
\sum_{r\in \bsz_N} \,\sum_{s=K+1}^{N+K-1}\, 
C_{\xi}(r,s)\dketm{r,s}~.
\end{eqnarray}

Let us try to solve the bootstrap equations 
\eqn{BO eq 1}, \eqn{BO eq 2}, assuming the Ansatz 
\begin{eqnarray}
\Psi_{\xi}(p,j)= f(p,j) S^{(\msc{c})}(p, j|\xi)~,~~~
C_{\xi}(r,s)= g(r,s) S^{(\msc{d})}(r, s|\xi)~,
\label{N=2 Liouville ansatz}
\end{eqnarray}
as in the previous examples.
First of all, the basic boundary state $\ket{B;O}$ is readily determined  
from the modular transformation 
formula \eqn{Liouville massless S 2};
\begin{eqnarray}
&& \ket{B;O} = \int_0^{\infty}dp'\, \sum_{j' \in \bsz_{2NK}}\,
\Psi_O(p',j')\dket{p',j'}+ 
\sum_{r'\in \bsz_{N}} \,\sum_{s'=K+1}^{N+K-1}\, 
C_O(r',s')\dketm{r',s'}~, 
\label{N=2 Liouville B O} \\
&& \Psi_O(p',j') = \frac{1}{\cQ}\left(\frac{2}{NK}\right)^{1/4}
\frac{\Gamma\left(\frac{1}{2}+\frac{j'}{2K}+i\frac{p'}{\cQ}\right)
\Gamma\left(\frac{1}{2}-\frac{j'}{2K}+i\frac{p'}{\cQ}\right)}
{\Gamma(i\cQ p') \Gamma\left(1+i\frac{2p'}{\cQ}\right)}~, 
\label{N=2 Liouville Psi O} \\
&& C_O(r',s') = \left(\frac{2}{N}\right)^{1/2} 
\sqrt{\sin\left(\frac{\pi (s'-K)}{N}\right)}~, 
\label{N=2 Liouville C O}
\end{eqnarray}
Then from the equations \eqn{modular data 0}, \eqn{BO eq 1} we can determine 
\begin{eqnarray}
&& f(p',j')\equiv  \frac{1}{\Psi_O(p',j')^*} =
\cQ \left(\frac{NK}{2}\right)^{1/4}\,
\frac{\Gamma(-i\cQ p')\Gamma\left(1-i\frac{2p'}{\cQ}\right)}
{\Gamma\left(\frac{1}{2}+\frac{j'}{2K}-i\frac{p'}{\cQ}\right)
\Gamma\left(\frac{1}{2}-\frac{j'}{2K}-i\frac{p'}{\cQ}\right)}~,~~~ \nn
&& g(r',s') \equiv \frac{1}{C_O(r',s')} 
= \left(\frac{N}{2}\right)^{1/2} 
\frac{1}{\sqrt{\sin\left(\frac{\pi (s'-K)}{N}\right)}}~.
\label{N=2 Liouville f g}
\end{eqnarray}
In the case of general graviton representations $|B;r\rangle$ one has
\begin{equation}
\Psi_{r}(p',j')=e^{-2\pi i \frac{rj'}{N}}\Psi_O(p',j'),
\hskip2mm C_r(r',s')=e^{-2\pi i \frac{r(s'+2Kr')}{N}}C_O(r',s')~,
\label{general graviton}
\end{equation}
from \eqn{N=2 Liouville ansatz}, \eqn{N=2 Liouville f g} and the modular 
transformation formula \eqn{Liouville massless S 2}.

As for the massless matter representations, 
there is a slight difficulty
in deriving  the corresponding Cardy states. 
This is because Ishibashi 
states are spanned by $|r,s\rangle\rangle_M$ with $1\le s 
\le N+2K-1$ while there appear the extra ``boundary terms" 
$\chim^{}(r,s=K)$ and $\chim^{}(r,s=N+K)$ in the R.H.S. 
of \eqn{Liouville massless S 1}.  We must thus combine
the characters so that the terms $\chim^{~}(r,s=K)$ 
and $\chim^{~}(r,s=N+K)$
cancel in their modular transforms.
The minimal combinations with such a property is given by 
a pair of representations
with $U(1)$ charges differing by an odd integer 
\begin{eqnarray}
&& \lb (r_1,s_1), \, (r_2,s_2)\rb \, \longleftrightarrow\, 
\chim^{}(r_1,s_1;\tau,z) + \chim^{}(r_2,s_2;\tau,z)~,\nn
&& (s_1+2Kr_1)-(s_2+2Kr_2) \in N(2\bz+1)~.
\label{class 3}
\end{eqnarray}
For such a pair of representations \eqn{class 3} we can solve \eqn{BO eq 1} 
in the form
\begin{eqnarray}
&&\Psi_{\lb (r_1,s_1),\,(r_2,s_2)\rb} (p,j)=f(p,j) 
\left(S^{(\msc{c})}(p,j| r_1,s_1)+S^{(\msc{c})}(p,j| r_2,s_2)
\right)~,~ \nn
&&
C_{\lb (r_1,s_1),\,(r_2,s_2)\rb} (r,s)=g(r,s) 
\left(S^{(\msc{d})}(r,s| r_1,s_1)+S^{(\msc{d})}(r,s| r_2,s_2)
\right)~,
\end{eqnarray}
where the modular coefficients $S^{(\msc{c})}(p,j| r_i,s_i)$,
$S^{(\msc{d})}(r,s| r_i,s_i)$ are read off from the formula
\eqn{Liouville massless S 1}.

In this way, we obtain three types of candidate Cardy states  
\begin{itemize}
 \item {\bf class 1} : Boundary states associated 
to the graviton representations 
    $\chig^{}(r;\tau,z)$, which we denote as $\ket{B;r}$ 
    ($r\in \bz_{N}$). Especially, 
     $\ket{B;r=0}\equiv \ket{B;O}$.
 \item {\bf class 2} : Boundary states associated 
to the massive representations 
    $\chi(p,j;\tau,z)$, which we denote as $\ket{B;p,j}$.
The parameters $p$, $j$ have to be constrained by the unitarity 
condition \eqn{unitarity bound class 2}.
\item {\bf class 3} : Boundary states associated to 
pairs of massless matter representations \\
    $\chim^{}(r_1,s_1;\tau,z)+\chim^{}(r_2,s_2;\tau,z)$
    \eqn{class 3},
    which we denote as $\ket{B; (r_1,s_1), (r_2,s_2)}$.
\end{itemize}

Of course this is not the whole story. We must check that these
candidate states in fact satisfy the Cardy 
condition \eqn{Cardy condition}. It is a non-trivial 
task to check whether  the overlaps
$\bra{B;\xi_1(\neq O)}e^{-\pi T H^{(c)}}\ket{B;\xi_2(\neq O)}$
can be rewritten in the form  \eqn{Cardy condition} with positive 
spectral densities.
We leave the analysis of these cylinder amplitudes in the next 
subsection, and present several comments here:

~

\noindent
{\bf 1.~}
Since  $\Psi_O(p,j=0)$ has a simple zero 
at $p=0$,  the solutions of \eqn{BO eq 1} have an ambiguity. 
We may add the term of the form  $\sim c \delta(p)$ to 
  the function $f(p, 0)$. This would lead to a subtlety, because 
  if allowing  such term, the overlap amplitudes of boundary states would
  be ill-defined (due to the product of delta functions).
  In the bosonic (and $\cN=1$) Liouville theory, 
  it is known from the bootstrap analyses for disk one-point 
  functions that no such delta function terms appear in the boundary 
  wave functions \cite{FZZ,ZZ,FH,ARS}. We believe they do not exist 
  in our $\cN=2$ case either and shall neglect this possibility 
 from here on.

~


\noindent
{\bf 2.}~ 
Our modular bootstrap approach determines 
only the absolute values of the boundary wave 
functions and there exist phase ambiguities 
in the solutions \eqn{N=2 Liouville Psi O}, 
\eqn{N=2 Liouville C O}. These phases may depend on the values of the cosmological constant.
In particular, determination of the phase factor for $\Psi_O(p,j)$ 
(or equivalently that of $f(p,j)$) 
is an important problem 
since it could reproduce the disk amplitudes of general vertex operators
by suitable analytic continuations.

As is well-known, 
the boundary wave function of a general Cardy state $\ket{B;\xi}$
may be interpreted as 
the disk one-point function by a relation such as
\begin{eqnarray}
\langle e^{\beta \phi(z,\bar{z})} \rangle_{\msc{disc}} 
\approx \frac{\Psi_{\xi}^*(p,0)}{|z-\bar{z}|^{2h_{\beta}}} ~, 
~~~(\beta \equiv \frac{{\cal Q}}{2}+ip)~,
\label{one point disc}
\end{eqnarray}
where $e^{\beta \phi}$ is the primary field of conformal weight 
$\dsp h_{\beta}=-\frac{\beta(\beta-{\cal Q})}{2}\equiv 
\frac{p^2}{2}+\frac{{\cal Q}^2}{8}$.

A possible way to determine the phase is to make use of the 
``reflection relation'' \cite{FZZ};
\begin{eqnarray}
\Psi_O(-p,j)=R(p,j)\Psi_O(p,j)~,
\label{reflection relation}
\end{eqnarray}
where $R(p,j)$ denotes the reflection amplitude 
(bulk two point function). 
In \cite{BF,AKRS} a solution for 
the reflection amplitude 
has been proposed based on the analysis parallel to \cite{Teschner3} 
for the model with the Liouville potential $\mu(S_++S_-)$, $\mu \in
\br_{>0}$\footnote
  {We should thank Y. Nakayama for drawing our attention to the paper 
   \cite{AKRS}. We should also thank P. Baseilhac for informing us of
   \cite{BF}.}. 
It is rewritten in our convention as 
\begin{eqnarray}
R(p, j)= \hat{\mu}^{-2i\cQ p} \, \frac{\Gamma(i\cQ p) 
\Gamma\left(1+i\frac{2p}{\cQ}\right)
\Gamma\left(\frac{1}{2}+\frac{j}{2K}-i\frac{p}{\cQ}\right)
\Gamma\left(\frac{1}{2}-\frac{j}{2K}-i\frac{p}{\cQ}\right)}
{\Gamma(-i\cQ p) 
\Gamma\left(1-i\frac{2p}{\cQ}\right)
\Gamma\left(\frac{1}{2}+\frac{j}{2K}+i\frac{p}{\cQ}\right)
\Gamma\left(\frac{1}{2}-\frac{j}{2K}+i\frac{p}{\cQ}\right)}~,
\label{reflection amplitude}
\end{eqnarray}
where $\hat{\mu}$ is the renormalized cosmological constant
proportional to $\mu$ whose precise value is not important 
here.

If we compare \eqn{N=2 Liouville Psi O} with the reflection amplitude \eqn{reflection
amplitude}, we find that they are completely consistent with the relation
(\ref{reflection relation})
provided $\Psi_O(p,j)$ is multiplied by an 
extra phase factor $\hmu^{i \cQ p}$ as
\begin{equation}
\Psi_O(p,j)=\hmu^{i \cQ p}\frac{1}{\cQ}\left(\frac{2}{NK}\right)^{1/4}
\frac{\Gamma\left(\frac{1}{2}+\frac{j}{2K}+i\frac{p}{\cQ}\right)
\Gamma\left(\frac{1}{2}-\frac{j}{2K}+i\frac{p}{\cQ}\right)}
{\Gamma(i\cQ p) \Gamma\left(1+i\frac{2p}{\cQ}\right)}~. 
\label{modified Psi_O}\end{equation} 
We note that this phase factor is what is expected based on the simple scaling
argument for the (fractional) 
number of insertions of bulk Liouville potential terms for the one-point 
function (\ref{one point disc}).
Thus we suggest that (\ref{modified Psi_O}) is in fact the correct expression
of the boundary wave function of ${\cal N}=2$ Liouville theory. 

~

\noindent
{\bf 3.}~
As is obvious from our construction, the class 1 boundary states
are regarded as generalizations of the (1,1)-type ZZ state in 
the bosonic Liouville theory \cite{ZZ}. Especially the wave function
$\Psi_O(p,j)$ has a simple zero at $p=0$. This fact implies 
that the corresponding $D$-brane is localized at the strong coupling
region $\phi\,\sim\, +\infty$ as discussed in \cite{KMS}.
Furthermore,
we find that $\Psi^*_O(p,0)$ has a simple pole at
$ip=\cQ/2$, corresponding to the one point function of $e^{\cQ \phi}$. 
Since the wave function for the quantum state $e^{\cQ \phi}\ket{0}$
behaves as $\psi_{\cQ}(\phi) \sim g_s^{-1}e^{\cQ \phi}\sim 
e^{\frac{\cQ}{2}\phi}$ \cite{Seiberg-L}, 
the existence of this pole is  a signal such that  the brane of $\ket{B;O}$ 
is really located  at $\phi \sim + \infty$.

~

\noindent
{\bf 4.}~
The class 2 and 3 states appear to describe analogues of the FZZT branes
\cite{FZZ,Teschner} which are extended along the Liouville direction.
We have a simple pole at $p=0$ this time, suggesting the 
``Neumann''  boundary condition along the Liouville direction. 
Poles of $\Psi^*_{\xi}(p,j)$ are determined  by the function
$f(p,j)^*$ which has the following form 
\begin{eqnarray}
&& f(p,j)^* \approx
\mu^{-\frac{\cQ}{2}ip-\frac{j}{2N}}
\bar{\mu}^{-\frac{\cQ}{2}ip+\frac{j}{2N}}
\frac{\Gamma(i\cQ p)\Gamma\left(1+i\frac{2p}{\cQ}\right)}
{\Gamma\left(\frac{1}{2}+\frac{j}{2K}+i\frac{p}{\cQ}\right)
\Gamma\left(\frac{1}{2}-\frac{j}{2K}+i\frac{p}{\cQ}\right)}~.
\label{f *}
\end{eqnarray}
Here we have also introduced an extra phase factor 
as in (\ref{modified Psi_O})
(with complex cosmological constants $\mu$ and $\bar{\mu}$).
\eqn{f *} implies that
the one point function \eqn{one point disc} has simple  
poles at $\dsp ip = -\frac{n}{\cQ}$, ($n\in \bz_{\geq 0}$).
In the perturbative approach to Liouville theory,
these poles could be interpreted as 
the results of an {\em integral\/} number of 
insertions of the boundary cosmological constant operators
$S^{(B)}_{\pm}$ which are defined like
the bulk operators \eqn{cosm chiral} 
but with a half Liouville momenta\footnote
   {Recently, a detailed study on the explicit forms of $S^{(B)}_{\pm}$
    has been given in \cite{AY}.}.
In fact, the location of 
these poles as well as the factor $\mu^{n/2} \bar{\mu}^{n/2}\equiv |\mu|^n$ 
fit nicely to the amplitude
$
\langle e^{\beta\phi} \, (S^{(B)}_+ S^{(B)}_-)^n \rangle_{0, \msc{disc}}
$ 
where the subscript ``0'' indicates the free field calculation.
Note that we need the same number of $S^{(B)}_+$ and $S^{(B)}_-$
insertions to obtain a non-vanishing result 
due to $U(1)$-charge conservation in free field calculation.   
These facts support our suggestion that the class 2 and class 3 states
describe the analogues of FZZT-branes.
Of course, these expectations should be confirmed  by the bootstrap
calculation for the disc one-point functions, which will 
clarify the precise interpretations of these boundary states 
based on the boundary Liouville interactions.

~

\noindent
{\bf 5.}~ Since $\ket{B;O}$ only includes the Ishibashi states 
in the NS sector, we cannot determine the R-sector Cardy states
based on the equation \eqn{BO eq 1}. We can instead  determine
them by means of the half-integral spectral flow, as we will 
do explicitly in the next section. The boundary wave functions 
for the R-sector determined this way have expected  forms 
including the modular coefficients of 
$\chi_*^{(\stNS)}(-1/\tau,z/\tau)$ rather than
$\chi_*^{(\sNS)}(-1/\tau,z/\tau)$.

~

\subsection{Cylinder Amplitudes and Cardy Condition}

~

Let us make an analysis on the overlaps (cylinder amplitudes) 
of the candidate boundary states to examine the Cardy condition.
The calculation is straightforward and we will see the relation
\eqn{Cardy condition} is satisfied in all the cases except for 
the overlaps between the class 3 states, that is,  of the type 
$\bra{B;(r_1,s_1),\,(r_2,s_2)}e^{-\pi T H^{(c)}} e^{i\pi z(J_0+\tJ_0)}
\ket{B;(r'_1,s'_1),\,(r'_2,s'_2)}$. 
It is rather non-trivial to check the Cardy condition in this case
and we will present its analyses in Appendix D.   
In this section we restrict ourselves to the class 1 and 2 boundary states.
We assume that the momentum $p$ labeling the class 2 states 
is real for the time being, and later discuss the cases of imaginary 
$p$ (bounded by the unitarity condition \eqn{unitarity bound class 2}).

First of all, the class 1 states have the following overlap amplitudes
\begin{eqnarray}
&&e^{\pi \frac{\hat{c}z^2}{T}}
\bra{B;r'} e^{-\pi T H^{(c)}} e^{i\pi z(J_0+\tJ_0)}
\ket{B;r} = \chig^{}(r-r';it,z')~, \\
&&e^{\pi \frac{\hat{c}z^2}{T}}
\bra{B;r'} e^{-\pi T H^{(c)}} e^{i\pi z(J_0+\tJ_0)}\ket{B;p,j} 
= \chi(p,j-2Kr';it,z')~, \\
&&e^{\pi \frac{\hat{c}z^2}{T}}
 \bra{B;r'} e^{-\pi T H^{(c)}} e^{i\pi z(J_0+\tJ_0)}
\ket{B;(r_1,s_1), (r_2,s_2)} \nn
&& \hspace{2cm} 
= \chim^{}(r_1-r', s_1;it,z') +\chim^{}(r_2-r', s_2;it,z')~, 
\end{eqnarray}
where the open string modulus $t$, $z'$ are defined in \eqn{BO eq 1},
\eqn{BO eq 2}.

We next calculate  the overlaps not including the class 1 states
\begin{eqnarray}
&&e^{\pi \frac{\hat{c}z^2}{T}} 
\bra{B;p_1,j_1}e^{-\pi T H^{(c)}}e^{i\pi z(J_0+\tJ_0)}
\ket{B;p_2,j_2} =
\int_0^{\infty}dp\, \left\lb 
\rho_1(p|p_1,p_2) \chi(p,j_2-j_1;it,z')
\right. \nn
&& \hspace{1.5cm} \left. + \rho_2(p|p_1,p_2)
\left\{ \chi(p,j_2-j_1+N;it,z') +\chi(p,j_2-j_1-N;it,z')
\right\} \right\rb~, \label{int rho}\\
&& \rho_1(p|p_1,p_2) = \int_0^{\infty}dp'\,
\frac{\cos(2\pi p p')}
{\sinh(\pi \cQ p')\sinh\left(\frac{2\pi p'}{\cQ}\right)}\,
\sum_{\ep_i=\pm 1}
\, \cosh\left(2\pi \left(
\frac{1}{\cQ}+i\ep_1p_1+i\ep_2p_2\right)p'\right)~, \nn
&& \label{rho 1}\\
&& \rho_2(p|p_1,p_2) = \frac{1}{2}\int_0^{\infty}dp'\,
\frac{\cos(2\pi p p')}
{\sinh(\pi \cQ p')\sinh\left(\frac{2\pi p'}{\cQ}\right)}\,
\sum_{\ep=\pm 1}
\, \cos\left(2\pi \left(p_1+\ep p_2\right)p'\right)~, 
\label{rho 2}
\end{eqnarray}
and also
\begin{eqnarray}
&&e^{\pi \frac{\hat{c}z^2}{T}} 
\bra{B;p_1,j_1}e^{-\pi T H^{(c)}}e^{i\pi z(J_0+\tJ_0)}
\ket{B;(r_1,s_1),(r_2,s_2)} =
\sum_{i=1,2}\, \int_0^{\infty}dp\, \nn
&& \hspace{5mm} \times
\left\lb \hat{\rho}_1(p|p_1,s_i) \chi(p,(s_i+2Kr_i)-j_1;it,z')
+ \hat{\rho}_2(p|p_1,s_i)
\chi(p,(s_i-N+2Kr_i)-j_1;it,z') \right\rb~, 
\label{int hat rho}\nn
&&  \\
&& \hat{\rho}_1(p|p_1,s) = \int_0^{\infty}dp'\,
\frac{\cos(2\pi p p')}
{\sinh(\pi \cQ p')\sinh\left(\frac{2\pi p'}{\cQ}\right)}\,
\sum_{\ep=\pm 1}
\, \cosh\left(2\pi \left(
\frac{N+K-s}{N}\frac{1}{\cQ}+i\ep p_1\right)p'\right)~, 
 \label{hat rho 1}\\
&& \hat{\rho}_2(p|p_1,s) = \int_0^{\infty}dp'\,
\frac{\cos(2\pi p p')}
{\sinh(\pi \cQ p')\sinh\left(\frac{2\pi p'}{\cQ}\right)}\,
\sum_{\ep=\pm 1}
\, \cosh\left(2\pi \left(
\frac{s-K}{N}\frac{1}{\cQ}+i\ep p_1\right)p'\right)~.
 \label{hat rho 2}
\end{eqnarray}
The spectral densities \eqn{rho 1}, \eqn{rho 2}, \eqn{hat rho 1} and
\eqn{hat rho 2} have similar forms as in the case of FZZT-branes. 
The momentum integral has a divergence 
at $p'=0$ as in the bosonic Liouville case. 
Such a divergence is not surprising,  
since the Liouville direction is non-compact and the class 2 and 3 
branes correspond to the Neumann boundary condition along this direction.
We note that the coefficients of $1/p'^2$ of the integrand 
are all positive and it is easy to check the positivity of 
the spectral functions.

After subtracting off the divergent piece which is independent of the
boundary states we find that the convergent part of
spectral densities 
are written in terms of the ``q-gamma functions'' 
defined in \cite{FZZ} (see \eqn{q-gamma});
\begin{eqnarray}
\rho_1(p|p_1,p_2) &\approx& \frac{i}{2\pi}\sum_{\ep_i=\pm 1}\,\ep_0\partial_{p}
\ln S_{\cQ/\sqrt{2}}\left(\frac{1}{\sqrt{2}}\left(\frac{\cQ}{2}
 +  i\ep_0 p +i\ep_1 p_1+ i\ep_2 p_2\right)\right) ~, 
\label{rho 1 2} \\
\rho_2(p|p_1,p_2) &\approx& \frac{i}{2\pi}\sum_{\ep_i=\pm 1}\,\partial_{p}
\ln S_{\cQ/\sqrt{2}}\left(\frac{1}{\sqrt{2}}\left(\frac{\cQ}{2}
+\frac{1}{\cQ}+ ip +i\ep_1 p_1+ i\ep_2 p_2\right)\right) ~,
\label{rho 2 2} \\
\hat{\rho}_1(p|s,p_1) &\approx& \frac{i}{2\pi}\sum_{\ep_i=\pm 1}\,\ep_0
\partial_{p}
\ln S_{\cQ/\sqrt{2}}\left(\frac{1}{\sqrt{2}}\left(\frac{\cQ}{2}
 + \frac{1}{\cQ}\frac{s-K}{N}+ i\ep_0 p +i\ep_1 p_1\right)\right) ~, 
\label{hat rho 1 2} \\
\hat{\rho}_2(p|s,p_1) &\approx& \frac{i}{2\pi}\sum_{\ep_i=\pm 1}\,
\ep_0\partial_{p}
\ln S_{\cQ/\sqrt{2}}\left(\frac{1}{\sqrt{2}}\left(\frac{\cQ}{2}
 + \frac{1}{\cQ}\frac{N+K-s}{N}+ i\ep_0 p +i\ep_1 p_1\right)\right) ~. 
\label{hat rho 2 2} 
\end{eqnarray}
The q-gamma function appears in the calculations of 
disc two-point functions (or the reflection 
coefficients) in the $\cN=0$ and $\cN=1$ Liouville theories,  
and the quantum mechanical relation between 
the two-point functions and the spectral densities has been proposed 
in \cite{ZZ}. It is an interesting problem to compute 
the disc two-point functions in the $\cN=2$ Liouville theory
and compare them with the spectral densities we have derived here.

We finally discuss  the class 2 states with 
pure imaginary $p$. Note that we allow only real values of 
momenta $p$ for Ishibashi states,
however, imaginary values of momenta $p$ possibly appear among Cardy states.
If one considers such a state, the integral over $p'$ in  
\eqn{rho 1}, \eqn{rho 2}, \eqn{hat rho 1}, \eqn{hat rho 2}
may generate additional divergences at 
$p'=\infty$. 
One may eliminate such a divergence 
by shifting the contour of $p$ integration
in (\ref{int rho}), (\ref{int hat rho}). 
When one shifts back the contour of $p$ to the real axis,  
one picks up additional contributions
from the poles of the q-gamma functions as discussed in \cite{Teschner}. 
These discrete terms correspond to the non-degenerate representations  
below the mass gap and appear with positive integer
coefficients. 
(Actually, we find the factor +1 per each pole.) 
However, they  do not always satisfy the  
unitarity bound \eqn{unitarity bound class 2}.
Although the complete classification of the class 2 states compatible 
with unitarity is an interesting and tractable problem,  
we here only point out the following fact: if $|p|<\cQ/4$,
the class 2 states with imaginary $p$ do not contain the discrete terms 
in arbitrary overlaps with them.
In these cases overlap amplitudes are obviously compatible with unitarity. 


In this way, we have established that all the class 1 states 
and class 2 states 
satisfying 
\begin{eqnarray}
\frac{p^2}{2} 
~ \left\{
\begin{array}{ll}
\dsp \geq  -\frac{(j_0-K)^2}{4NK}~,  & \dsp ~~
\mbox{if}~ (j_0-K)^2 < \frac{K^2}{4}\\
\dsp  > -\frac{K}{16N} ~,& \dsp ~~ \mbox{if}~  (j_0-K)^2 \geq  \frac{K^2}{4}
\end{array}
\right. ~~~,  ~~~ 0\leq j_0 < 2K~,~~ 
j_0\equiv j ~ (\mod\, 2K)~
\label{bound class 2}
\end{eqnarray}
are compatible with  
the Cardy condition \eqn{Cardy condition}
and hence describe consistent $D$-branes in the 
general $\cN=2$ Liouville theory.

~

\subsection{Summary of Cardy States in $\cN=2$ Liouville Theory}

~

To close this section let us present a summary of the (NS) Cardy states 
in $\cN=2$ Liouville theory we propose for the convenience of readers:

\noindent
{\bf general form :}
\begin{eqnarray}
&& \ket{B;\xi} = \int_0^{\infty}dp'\, \sum_{j \in \bsz_{2NK}}\,
\Psi_{\xi}(p',j')\dket{p',j'}+ 
\sum_{r\in \bsz_N} \,\sum_{s'=K+1}^{N+K-1}\, 
C_{\xi}(r',s')\dketm{r',s'}~,
\end{eqnarray}
where the Ishibashi states $\dket{p',j'}$, $\dketm{r',s'}$
are defined in \eqn{Ishibashi N=2 Liouville}, \eqn{Ishibashi N=2
Liouville OC}. The boundary wave function $\Psi_{\xi}(p',j')$ 
could include the scaling factor proportional to 
$\mu^{-\frac{\cQ}{2}ip'-\frac{j'}{2N}}
\bar{\mu}^{-\frac{\cQ}{2}ip'+\frac{j'}{2N}}$ (irrespective of $\xi$),
which is just a phase factor for real $p'$.
\begin{itemize}
\item
 {\bf class 1 states} :  ~~~ $\ket{B;r}$ ~~ ($r\in \bz_{N}$)
\begin{eqnarray}
&& \Psi_r(p',j') = \frac{1}{\cQ}\left(\frac{2}{NK}\right)^{1/4}
e^{-2\pi i \frac{rj'}{N}}
\frac{\Gamma\left(\frac{1}{2}+\frac{j'}{2K}+i\frac{p'}{\cQ}\right)
\Gamma\left(\frac{1}{2}-\frac{j'}{2K}+i\frac{p'}{\cQ}\right)}
{\Gamma(i\cQ p') \Gamma\left(1+i\frac{2p'}{\cQ}\right)} ~, 
\nn
&& C_r(r',s') = \left(\frac{2}{N}\right)^{1/2} 
e^{-2\pi i \frac{r(s'+2Kr')}{N}}
\sqrt{\sin\left(\frac{\pi (s'-K)}{N}\right)}~.
\end{eqnarray} 
\item
{\bf class 2 states} :  ~~~ $\ket{B;p,j}$ 
\begin{eqnarray}
\frac{p^2}{2} \geq - \frac{(j_0-K)^2}{4NK}~,~~
 j_0 \equiv j ~,~  (\mod\, 2K)~, ~~  0\leq j_0 \leq 2K-1~.
\end{eqnarray}
\begin{eqnarray}
&& \Psi_{p,j}(p',j') =
\cQ \left(\frac{2}{NK}\right)^{1/4}\, e^{-2\pi i \frac{jj'}{2NK}} 
\cos(2\pi pp')\, 
\frac{\Gamma(-i\cQ p')\Gamma\left(1-i\frac{2p'}{\cQ}\right)}
{\Gamma\left(\frac{1}{2}+\frac{j'}{2K}-i\frac{p'}{\cQ}\right)
\Gamma\left(\frac{1}{2}-\frac{j'}{2K}-i\frac{p'}{\cQ}\right)}~,~~~ \nn
&& C_{p,j}(r',s') =0 ~.
\end{eqnarray}
\item
{\bf class 3 states} :  ~~~ $\ket{B; (r_1,s_1), (r_2,s_2)}$ 
\begin{eqnarray}
&& r_i\in \bz_N~,~~~ 1\leq s_i \leq N+2K-1 ~,~~~
(s_1+2Kr_1)-(s_2+2Kr_2) \in N(2\bz+1)~. \nn
&&
\end{eqnarray}
\begin{eqnarray}
&& \Psi_{(r_1,s_1), (r_2,s_2)}(p',j') =
\cQ \left(\frac{NK}{2}\right)^{1/4}\, 
\left(S^{(\msc{c})}(p',j'|r_1,s_1)+ S^{(\msc{c})}(p',j'|r_2,s_2)\right)\,
\nn
&& \hspace{4cm} \times 
\frac{\Gamma(-i\cQ p')\Gamma\left(1-i\frac{2p'}{\cQ}\right)}
{\Gamma\left(\frac{1}{2}+\frac{j'}{2K}-i\frac{p'}{\cQ}\right)
\Gamma\left(\frac{1}{2}-\frac{j'}{2K}-i\frac{p'}{\cQ}\right)}~,~~~ \nn
&& C_{(r_1,s_1), (r_2,s_2)}(r',s') =   
 \left(\frac{N}{2}\right)^{1/2} 
\frac{1}{\sqrt{\sin\left(\frac{\pi (s'-K)}{N}\right)}}\,
\left(
S^{(\msc{d})}(r',s'|r_1,s_1)+ S^{(\msc{d})}(r',s'|r_2,s_2)
\right)~, \nn
&& S^{(\msc{c})}(p',j'|r,s) 
= \frac{1}{\sqrt{2NK}}e^{-2\pi i \frac{(s+2Kr)j'}{2NK}}\,
\frac{\cosh\left(2\pi \frac{N+K-s}{N}\frac{p'}{\cQ}\right)
+e^{i\frac{\pi j'}{K}} \cosh\left(2\pi \frac{s-K}{N}\frac{p'}{\cQ}\right) }
{2\left|\cosh \, \pi \left(\frac{p'}{\cQ}+i\frac{j'}{2K}\right)\right|^2}
~, \nn 
&& S^{(\msc{d})}(r',s'|r,s)
= \frac{i}{N} e^{-2\pi i \frac{(s+2Kr)(s'+2Kr')-(s-K)(s'-K)}{2NK}}~.
\end{eqnarray}
\end{itemize}

~


\section{Coupling with $\cN=2$ Minimal Models and 
Calabi-Yau Singularities}

~

As an application of our previous analysis let us consider 
the superstring vacua of the form;
\begin{eqnarray}
&& \br^{d-1,1} \times \left\lb \mbox{$\cN=2$ minimal model of level $k$ 
   ($\hat{c}=k/(k+2)$)}\right\rb \nn
&& \hspace{3cm}
\times 
\left\lb \mbox{$\cN=2$ Liouville $(\hat{c}=1+(2K)/N)$}\right\rb~, 
\label{coupled system}
\end{eqnarray}
with the criticality condition 
\begin{eqnarray}
 \frac{d}{2}+ \frac{k}{k+2} + \left(1+\frac{2K}{N}\right)= 5~.
\label{criticality}
\end{eqnarray}
Precisely speaking, we have to make an orbifoldization of this theory 
and project onto sectors with integral total $U(1)$-charges.
This procedure is often called the ``GSO projection''.\footnote
    {After projection to integral $U(1)$ charges, 
     we must further impose the usual 
     GSO projection  
     with respect to the spin structures, 
      which leads to, for instance, $Q_{\msc{total}} \in 2\bz+1$
      for the NS sector.} 
These superconformal systems are believed to  describe 
the type II string theories compactified on 
the Calabi-Yau $n$-folds ($n\equiv (10-d)/2$)
with isolated A-D-E singularities when one uses
modular invariants of the $\cN=2$ minimal model of the corresponding 
A-D-E type \cite{ABKS,GKP,ES1,NN}. 
For simplicity we shall work only with the A-type modular invariants, which correspond 
to the $A_{k+1}$-type Calabi-Yau singularity locally expressed as 
$X^{k+2}+ z_1^2+\cdots +z_n^2=0$ (for $CY_n$). 
The main  purpose of this section is to clarify how we can 
identify the vanishing cycles of these singularities 
with appropriate Cardy states. 
Earlier studies on this subject has been given in {\em e.g.}
\cite{Lerche,LLS,ES2}.

The criticality condition \eqn{criticality} leads to the following values of $N,K$
\begin{itemize}
 \item $d=6$ : $N=k+2$, $K=1$.
 \item $d=4$ : 
\begin{list}%
 {$\cdot$} 
 {} 
 \item  $k\in 2\bz_{\geq 0}$ : $N=k+2$, $K=(k+4)/2$, 
 \item $k\in 2\bz_{\geq 0}+1$ : $N=2(k+2)$, $K=k+4$. 
\end{list}
 \item $d=2$ : $N=k+2$, $K=k+3$.
\end{itemize}

~

\subsection{BPS $D$-branes Wrapped Around Vanishing Cycles}

~

Now, we start the analysis on the BPS $D$-branes in these
superconformal systems.
We again concentrate on the A-type boundary states which 
describe the special Lagrangian cycles.. 

First of all, the Cardy states in the minimal model are well-known
(see, for instance, \cite{RS}).
Let $\dket{\ell,m}^{(\sNS)}$
($\dket{\ell,m}^{(\sR)}$) be 
the Ishibashi states in the NS (R) sector
characterized by
\begin{eqnarray}
&& {}^{(\sNS)}\dbra{\ell,m} e^{-\pi T H^{(c)}}
e^{i\pi z(J_0+\tJ_0)}
 \dket{\ell',m'}^{(\sNS)} 
= \left(\delta_{\ell,\ell'}\delta_{m,m'}
+\delta_{\ell,k-\ell'}\delta_{m,m'+k+2}
\right) \, \ch{(\sNS)}{\ell,m}(iT,z)~, \nn
&&  {}^{(\sR)}\dbra{\ell,m} e^{-\pi T H^{(c)}} 
e^{i\pi z(J_0+\tJ_0)}
\dket{\ell',m'}^{(\sR)} 
= \left(\delta_{\ell,\ell'}\delta_{m,m'}
+\delta_{\ell,k-\ell'}\delta_{m,m'+k+2}
\right) \, \ch{(\sR)}{\ell,m}(iT,z)~, \nn
&& 
\end{eqnarray}
where $\ch{(\sNS)}{\ell,m}(\tau,z)$ ($\ch{(\sR)}{\ell,m}(\tau,z)$) 
denotes the NS (R) character of the $\cN=2$
minimal model with the primary field with
$\dsp h=\frac{\ell(\ell+2)-m^2}{4(k+2)}$, $\dsp Q=\frac{m}{k+2}$~
($\dsp h=\frac{\ell(\ell+2)-m^2}{4(k+2)}+\frac{1}{8}$, 
$\dsp Q= \frac{m}{k+2}\pm \frac{1}{2}$). (See Appendix A.)
Note that the character identities 
\begin{eqnarray}
&& \ch{(\sigma)}{k-\ell,m+k+2}(\tau,z)= \ch{(\sigma)}{\ell,m}(\tau,z)
~,~~(\sigma = \NS,\, \R)~,\nn
&& \ch{(\sigma)}{k-\ell,m+k+2}(\tau,z)= -\ch{(\sigma)}{\ell,m}(\tau,z)
~,~~(\sigma = \tNS,\, \tR)~,
\end{eqnarray}
imply the field identification 
\begin{eqnarray}
\dket{k-\ell,m+k+2}^{(\sigma)}=\dket{\ell,m}^{(\sigma)}~.
\label{field identification}
\end{eqnarray}
We can thus restrict ourselves 
to the range $0\leq \ell \leq k$,  $m\in \bz_{k+2}$
without loss of generality. 
It is also convenient to set $\dket{\ell,m}^{(\sNS)}=0$
($\dket{\ell,m}^{(\sR)}=0$), if $\ell+m \in 2\bz+1$ ($\ell+m \in 2\bz$).
Then, the Cardy states are expressed as follows
($\sigma=\NS, ~ \mbox{or} ~ \R~$);
\begin{eqnarray}
&&\ket{L,M}^{(\sigma)} = \sum_{\ell=0}^k\,\sum_{m\in \bsz_{k+2}}\,
C_{L,M}(\ell,m)\dket{\ell,m}^{(\sigma)}~, ~~~
 C_{L,M}(\ell,m) = \frac{S^{L,M}_{\ell,m}}{\sqrt{S^{0,0}_{\ell,m}}}~,
\nn
&& L+M\in 2\bz~~ (\mbox{for}~ \sigma=\NS)~,~~~
   L+M\in 2\bz+1~~ (\mbox{for}~ \sigma=\R)~,
\label{minimal Cardy states}
\end{eqnarray}
where $S^{\ell,m}_{\ell',m'}$ is the modular coefficients of 
$\ch{(\sNS)}{\ell,m}(\tau,z)$, given explicitly by
\begin{eqnarray}
S^{\ell,m}_{\ell',m'} = 2 \cdot
\sqrt{\frac{2}{k+2}}\sin\left(\frac{\pi(\ell+1)(\ell'+1)}{k+2}\right)
\cdot \frac{1}{\sqrt{2(k+2)}}e^{2\pi i \frac{mm'}{2(k+2)}} ~.
\label{minimal S}
\end{eqnarray}
(The overall factor 2 is due to the choice of range $m\in \bz_{k+2}$ 
instead of $\bz_{2(k+2)}$.)

All the quantities in R-sector are generated by the 1/2-spectral 
flow $U_{1/2}$ ($\eta=1/2$ in 
\eqn{spectral flow}) from those of NS-sector. 
Then, we find 
\begin{eqnarray}
\dket{\ell,m}^{(\sR)}=U_{1/2}\dket{\ell,m+1}^{(\sNS)}~,~~~
\ket{L,M}^{(\sR)} = e^{-i\pi \frac{M}{k+2}}U_{1/2}
e^{-i\pi \frac{1}{2}(J_0+\tJ_0)} \ket{L,M+1}^{(\sNS)}~.
\label{minimal NS R}
\end{eqnarray}



Let us next consider the Liouville sector. 
Our aim is to identify the BPS $D$-branes wrapped around 
the vanishing cycles. As we already discussed, the class 1 Cardy states
$\ket{B;r}$ correspond to the $D$-branes localized in
the strong coupling region $\phi\,\sim\, +\infty$, 
which is near the isolated singularity in the Calabi-Yau space
in this context. Furthermore, the overlaps among the class 1 states 
only contains the open strings of discrete spectrum, implying 
compact world-volumes of $D$-branes. 
The class 2 and 3 states do not have such a property.
From these observations, 
it is plausible to guess that the class 1 states
capture the open string excitations moving near the singularity and 
characterize  the geometry of the singular Calabi-Yau space.
Therefore, we propose that\\

\noindent
{\bf Supersymmetric boundary states 
describing  the vanishing cycles are given by 
the class 1 states of ${\cal N}=2$ Liouville theory.} \\

Using the 1/2-spectral flow,
the extended characters for the graviton representation \eqn{Liouville
massless 2} for other spin structures are given by  
\begin{eqnarray}
&& \chig^{(\stNS)}(r;\tau,z)= e^{-i\pi \frac{2Kr}{N}}\, 
\chig^{(\sNS)}\left(r;\tau,z+\frac{1}{2}\right)~, ~~(r\in \bz_{N})
~, \nn
&&\chig^{(\sR)}(r;\tau,z)= q^{\hat{c}/8}y^{\hat{c}/2}\,  
\chig^{(\sNS)}\left(r-\frac{1}{2};\tau,z+\frac{\tau}{2}\right)~, 
~~(r\in \frac{1}{2}+\bz_{N})~, \nn
&&\chig^{(\stR)}(r;\tau,z)=  e^{-i\pi \frac{K(2r-1)}{N}}
q^{\hat{c}/8}y^{\hat{c}/2}\,  
\chig^{(\sNS)}\left(r-\frac{1}{2};\tau,z+\frac{\tau}{2}+\frac{1}{2}\right)~,
~~(r\in \frac{1}{2}+\bz_{N})~,\nn
&& \label{Liouville massless others}
\end{eqnarray}
and we define 
\begin{eqnarray}
&& \dket{p,j}^{(\sR)}= U_{1/2}\dket{p,j-K}^{(\sNS)}~,~~~
   \dketm{r,s}^{(\sR)} = U_{1/2}\dketm{r-1/2,s}^{(\sNS)}~, 
~~(r\in \frac{1}{2}+\bz_{N}) \nn
&& \ket{B;r}^{(\sR)}= e^{-i\pi \frac{K(2r-1)}{N}}U_{1/2}
 e^{-\frac{i\pi}{2}(J_0+\tJ_0)} \ket{B;r-1/2}^{(\sNS)}~,~~
(r\in \frac{1}{2}+\bz_{N})~,
\label{Liouville NS R}
\end{eqnarray}
similarly to  \eqn{minimal NS R}.



Now, taking account of the GSO projection, 
the proposed boundary states have the form
\begin{eqnarray}
\sqrt{N} \, P_{\msc{GSO}} \left(\ket{L,M}^{(\sigma)}\otimes 
\ket{B;r}^{(\sigma)}\right)~, 
\label{total Cardy 0}
\end{eqnarray}
where $P_{\msc{GSO}}$ stands for the GSO projection operator
imposing the integrality of total $U(1)$-charge in the closed string 
Hilbert space. $\sqrt{N}$ is a normalization constant fixed 
by the Cardy condition as we shall see below.

$P_{\msc{GSO}}$ restricts the choice of the tensor products of Ishibashi states 
to;
\begin{itemize}
\item NS-sector : 
\begin{eqnarray}
&& \dket{\ell, m}^{(\sNS)}\otimes \dketm{r,s}^{(\sNS)}~,~~ 
\frac{m}{k+2} + \frac{s+2Kr}{N} \in \bz~, ~~~(r\in \bz_N)~,  \nn
&& \dket{\ell, m}^{(\sNS)} \otimes  \dket{p,j}^{(\sNS)}~,~~ 
\frac{m}{k+2}+\frac{j}{N} \in \bz ~.
\end{eqnarray}
\item R-sector : 
\begin{eqnarray}
&& \dket{\ell, m}^{(\sR)}\otimes \dketm{r,s}^{(\sR)}~,~~ 
\frac{m}{k+2} + \frac{s+2Kr}{N} \in \bz~,~~~(r\in \frac{1}{2}+\bz_N)~,
~~ \mbox{for $d=2,6$}~, \nn
&& \hspace{4.3cm} 
\frac{m}{k+2} + \frac{s+2Kr}{N} -\frac{1}{2} \in \bz~,~~~
(r\in  \frac{1}{2}+\bz_N)~,
~~ \mbox{for $d=4$}~, \nn
&&\dket{\ell, m}^{(\sR)} \otimes  \dket{p,j}^{(\sR)}~,~~ 
\frac{m}{k+2} + \frac{j}{N} \in \bz~,
~~ \mbox{for $d=2,6$}~, \nn
&& \hspace{4.8cm}
\frac{m}{k+2} + \frac{j}{N} -\frac{1}{2} \in \bz~,
~~ \mbox{for $d=4$}~, 
\end{eqnarray} 
\end{itemize}
From (\ref{general graviton}) it is easy to find that \eqn{total Cardy 0} depends on 
the parameters $M$, $r$ only through the combination $M+2r$.
We can thus simply set $r=0$ ($r=1/2$) for the NS (R) sector
(namely, $\ket{B;O}$ itself for the NS-sector) without loss of generality.
In this way we consider the boundary states 
\begin{eqnarray}
&&\ket{B;L,M}^{(\sNS)} = \sqrt{N}\, P_{\msc{GSO}}
\left(\ket{L,M}^{(\sNS)}\otimes 
\ket{B;r=0}^{(\sNS)}\right)~,~\nn
&&\ket{B;L,M}^{(\sR)} = \sqrt{N}\, P_{\msc{GSO}}
\left(\ket{L,M}^{(\sR)}\otimes 
\ket{B;r=1/2}^{(\sR)}\right)~,
\label{total Cardy}
\end{eqnarray}
and set
\begin{eqnarray}
\ket{B;L,M,\pm} = \ket{B;L,M}^{(\sNS)}\pm \ket{B;L,M-1}^{(\sR)}~,~~~
(L+M\in 2\bz)~.
\end{eqnarray}
Obviously $\ket{B;L,M,+}$ and $\ket{B;L,M,-}$ can be 
interpreted as the brane and anti-brane respectively.
We also note 
\begin{eqnarray}
\hskip-3mm \ket{B;k-L,M+k+2}^{(\sNS)}= \ket{B;L,M}^{(\sNS)}~,~
\ket{B;k-L,M+k+2}^{(\sR)}= -\ket{B;L,M}^{(\sR)}~,
\end{eqnarray}
and thus 
\begin{eqnarray}
\ket{B;k-L,M+k+2,\pm}= \ket{B;L,M,\mp}~. 
\label{brane anti-brane}
\end{eqnarray}
Therefore, it is enough to restrict ourselves to the range $\dsp L=0,1,\ldots, 
\left\lb \frac{k}{2}\right\rb$, $M\in \bz_{2N}$, as long as 
we consider only the branes (or the anti-branes).

Let us analyze the overlap amplitudes of the total boundary states
\eqn{total Cardy}. We first consider the NS sector.
Presence of $P_{\msc{GSO}}$ is the only non-trivial point 
in this  calculation.
The easiest way to handle it is to insert 
\begin{eqnarray}
\delta^{(N)}\left(\frac{N}{k+2}m+
\beta\right) \equiv \frac{1}{N}\,
\sum_{r\in \bsz_{N}}\,
e^{- 2i \pi r\left(\frac{1}{k+2}m+
\frac{1}{N}\beta\right)}
~, ~~~ (\beta =j ~\mbox{or} ~ s+2Kr)~,
\label{insertion NS}
\end{eqnarray}
into the amplitudes.
We obtain after a little calculation ($\dsp \hat{c}=5-\frac{d}{2}$);
\begin{eqnarray}
&& e^{\pi\frac{\hat{c}z^2}{T}}\,
{}^{(\sNS)}\bra{L_1,M_1} e^{-\pi T H^{(c)}} e^{i\pi z(J_0+\tJ_0)}
\ket{L_2,M_2}^{(\sNS)} \nn
&& \hspace{2cm }= 
\sum_{L=0}^{k}\, \sum_{r\in \bsz_{N}}\,
N_{L_1,L_2}^L\, \ch{(\sNS)}{L,M_2-M_1-2r}(it,z')\chig^{(\sNS)}(r;it,z')~,
\label{overlap total}
\end{eqnarray}
where $N_{L_1,L_2}^L$ are the familiar fusion coefficients
of $SU(2)_k$;
\begin{eqnarray}
N_{L_1,L_2}^L=
\left\{
\begin{array}{ll}
 1& ~~ |L_1-L_2|\leq L \leq \min\lb L_1+L_2,\, 2k-L_1-L_2\rb~,~ 
       L \equiv |L_1-L_2|~ (\mod\, 2)\\
 0& ~~ \mbox{otherwise}
\end{array}
\right.
\end{eqnarray}
Normalization factor $\sqrt{N}$ of the boundary state 
has been chosen so that the factor $1$ appears
in front of the R.H.S of (\ref{overlap total}).

The overlap \eqn{overlap total} implies that the open strings
have integral $U(1)$-charges, if and only if $M_1 \equiv M_2~(\mod\, k+2)$.
Careful examination of other spin structures leads to the
result: {\em the brane configurations are supersymmetric if and only if 
$M_1 \equiv M_2~(\mod\,\, 2(k+2))$.} 
(The cases of $M_1\equiv M_2 +(k+2)~(\mod\, 2(k+2))$ 
can be reinterpreted  as 
the $D$-$\bar{D}$ systems.)
Similar observation has been given in \cite{HNS}
in the context of supersymmetric $SU(2)_k$ WZW model in the $d=6$ case.
When $M_1 \equiv M_2 ~ (\mod\, 2(k+2))$ holds, 
the cancellation of cylinder amplitude due to 
the space-time SUSY takes place
\begin{eqnarray}
&& \sum_{\sigma}\sum_{L=0}^k\sum_{r\in \bsz_{N}
+ \al(\sigma)}
\,\ep(\sigma)
\left(\frac{\theta_{\lb \sigma \rb}(\tau,0)}{\eta(\tau)}\right)^{\frac{d-2}{2}}
\, N_{L_1,L_2}^L \,
\, \ch{(\sigma)}{L,-2r}(it,0)\chig^{(\sigma)}(r;it,0)
=0~. 
\label{SUSY cancellation}  
\end{eqnarray}
Here $(\theta_{\lb \sigma \rb}(\tau,0)/\eta(\tau))^{(d-2)/2}$ part comes
from the transverse space-time $\br^{d-2}$ and 
we set $\ep=+1$ for $\sigma=\NS,\tR$, $\ep=-1$ for $\sigma=\tNS,\R$,
$\al=0$ for $\sigma=\NS,\tNS$, $\al=1/2$ for $\sigma=\R,\tR$,
and $\theta_{\lb \sNS \rb}=\theta_3$, $\theta_{\lb \stNS \rb}=\theta_4$,
$\theta_{\lb \sR \rb}=\theta_2$, $\theta_{\lb \stR \rb}=i\theta_1$.
The above relation  can be proven as a special 
case of SUSY cancellation of the $\hat{c}=4$ extended characters
shown in \cite{HS}. Thus the self-overlaps of 
$\ket{B;L,M,\pm}$ are supersymmetric for arbitrary $L$, $M$ and
we conclude that they are indeed BPS $D(\bar{D})$-branes.

Now let us identify the boundary states $\ket{B;L,M,\pm}$
with the vanishing cycles in a singular Calabi-Yau manifold. 
In the case of A-type singularity vanishing cycles can be represented by line segments 
connecting pairs of roots of $X^{k+2}=\mu$ in the complex
$X$-plane (see \cite{ES2}, for instance). Here the deformation
parameter $\mu$ is identified with the 
cosmological constant associated with the operator $S_+$
\eqn{cosm chiral}.  We parameterize pairs of roots as  
\begin{eqnarray}
&& X_{L,M,+} = \mu^{1/(k+2)} e^{i\pi (M+L)/(k+2)}~,~~
 X_{L,M,-} = \mu^{1/(k+2)} e^{i\pi (M-L-2)/(k+2)}~, \\
&& L=0,1,\ldots,\left\lb \frac{k}{2}\right\rb~, ~ M\in \bz_{2(k+2)}~,~
L+M \in 2\bz~, \nn
\end{eqnarray}
and denote the corresponding cycle as $\gamma_{L,M}$. 
As has been often pointed out in the literature (see, \cite{Lerche,ES2} for
instance), parameters $L$, $M$ of the vanishing cycles 
are also identified as quantum numbers $L$ and $M$ of the minimal model
themselves.
We thus claim the correspondence
\begin{eqnarray}
  \ket{B;L,M,\pm}~\longleftrightarrow~ 
   \mbox{$D$ ($\bar{D}$) brane wrapped around } \gamma_{L,M}~.
\label{correspondence}
\end{eqnarray}
We here emphasize that the description of the Liouville sector of the boundary states
$\ket{B;L,M,\pm}$ is the essential new ingredient of this work.

As an important consistency check let us 
evaluate the open string Witten indices, which are identified
as the intersection numbers of vanishing cycles.
The evaluation is almost parallel to that of the overlap \eqn{overlap total},
but in place of \eqn{insertion NS} we have to insert a factor
\begin{eqnarray}
&& \delta^{(2N)}\left(\frac{N}{k+2}
   m+  \beta- \frac{N\ep}{2}\right)
- \delta^{(2N)}\left(\frac{N}{k+2}
   m+ \beta+N -\frac{N\ep}{2}\right) \nn
&& \hspace{2cm}
\equiv
\frac{1}{N}\,
\sum_{r\in \frac{1}{2}+\bsz_{N}}\,
e^{- 2i \pi r\left(\frac{1}{k+2}m+
\frac{1}{N}\beta\right)+ i\pi \ep r}~, 
~~~ (\beta =j ~\mbox{or} ~ s+2Kr)~,
\label{insertion R}
\end{eqnarray}
where $\ep=0$ for $d=2,6$ and $\ep = 1$ for $d=4$.
The relative minus sign in the left-hand-side of this equation
is due to the insertion of $e^{\frac{i\pi}{2}(F+\tF)}$,
where $F$ and $\tF$ denote the world-sheet fermion number 
operators.
Open string Witten indices are thus evaluated as 
\begin{eqnarray}
&& I(L_1,M_1|L_2,M_2) \equiv {}^{(\sR)}\bra{B;L_1,M_1,+}e^{-\pi T H^{(c)}}
e^{\frac{i\pi}{2}(F+\tF)} \ket{B;L_2,M_2,+}^{(\sR)} \nn
&& \hspace{1.5cm}
= - \sum_{L=0}^k\,\sum_{r\in \frac{1}{2}+\bsz_{N}}\,
e^{i\pi \ep r}N^{L}_{L_1,L_2}\,
\ch{(\stR)}{L,M_2-M_1-2r}(it,0) \chig^{(\stR)}(r;it,0)~.
\label{Witten index 1}
\end{eqnarray}
To proceed further, we recall the Witten index for the minimal model;
\begin{eqnarray}
\ch{(\stR)}{\ell,m}(\tau,0) = \delta^{(2(k+2))}(m-(\ell+1))
 -  \delta^{(2(k+2))}(m+(\ell+1))~.
\label{WI minimal}
\end{eqnarray}
On the other hand in the Liouville sector we have a formula
\begin{eqnarray}
\chig^{(\stR)}(r;\tau,0) = \delta^{(N)}(r-1/2) -\delta^{(N)}(r+1/2) ~,
\label{WI Liouville} 
\end{eqnarray}
derived from \eqn{Liouville massless 2} and 
the definition of $\tR$-character \eqn{Liouville massless
others}.\footnote
    {The easiest way to derive \eqn{WI Liouville} is to recall 
     the character identity 
\begin{eqnarray}
\chig^{(\stR)}(r;\tau,z) &=& q^{-\frac{K}{4N}}
\Th{2Kr}{NK}\left(\tau,\frac{2z}{N}\right)\frac{i\th_1(\tau,z)}{\eta(\tau)^3}
+ \chim^{(\stR)}(r,N;\tau,z) -\chim^{(\stR)}(r-1,2K;\tau,z)~, \nn
 \chim^{(\stR)}(r,s;\tau,z) &\equiv &e^{-i\pi \frac{s+K(2r-1)}{N}}
q^{\frac{\hat{c}}{8}}y^{\frac{\hat{c}}{2}}\,
\chim^{(\sNS)}\left(r-\frac{1}{2},s;\tau,z+\frac{\tau}{2}+\frac{1}{2}\right)
\nn
&=& \sum_{m\in \bsz}\,\frac{\left(yq^{N\left(m+\frac{2r+1}{2N}\right)}
\right)^{\frac{s-K}{N}}}{1-yq^{N\left(m+\frac{2r+1}{2N}\right)}}\,
y^{2K\left(m+\frac{2r+1}{2N}\right)} q^{NK\left(m+\frac{2r+1}{2N}\right)^2}\,
\frac{i\th_1(\tau,z)}{\eta(\tau)^3}~, \nonumber
\end{eqnarray}
which is shown by taking the $1/2$-spectral flow of 
\eqn{Liouville character relation}.
}
We finally obtain 
\begin{eqnarray}
&&I(L_1,M_1|L_2,M_2)
=\sum_{L=0}^k\,\sum_{a_i=\pm 1}\, \mbox{sgn}(a_1) \mbox{sgn}(a_2)
e^{\frac{i\pi}{2} \ep a_2} N^{L}_{L_1,L_2} \delta^{\left(2(k+2)\right)}
(M_1-M_2-a_1(L+1)+a_2)~. \nn
&&
\label{WI total}
\end{eqnarray}
This is the desired result which reproduce the correct intersection numbers
of vanishing cycles $\gamma_{L,M+1}$. 
In particular, we find for the ``fundamental cycles'' $\gamma_{0,M+1}$,
\begin{eqnarray}
&& I(0,M_1|0,M_2) = 2 \delta^{\left(2(k+2)\right)} (M_1-M_2)
- \delta^{\left(2(k+2)\right)} (M_1-M_2-2) -
 \delta^{\left(2(k+2)\right)} (M_1-M_2+2)~,\nn
&&
\label{int 6,2}\end{eqnarray} 
for $d=2,6$ and 
\begin{eqnarray}
&& I(0,M_1|0,M_2) = 
\delta^{\left(2(k+2)\right)} (M_1-M_2-2) 
-\delta^{\left(2(k+2)\right)} (M_1-M_2+2)~,
\label{int 4}\end{eqnarray} 
for $d=4$.

Note that (\ref{int 6,2}) agrees with the $A_N$ type extended
Cartan matrix while
(\ref{int 4}) is anti-symmetric in $M_1$ and $M_2$ as expected 
for Langrangian 3-cycles.

~

\section{$D$-branes in Type 0 String Vacua Based on the $\cN=2$ Liouville 
with $\hat{c}=5$}

~

Let us next discuss the type 0 string vacua constructed 
from the $\cN=2$ Liouville theory, motivated by the recent studies
on the duality between matrix models and non-critical strings, 
and unstable $D$-branes 
\cite{MV,KMS,MTV,Martinec,AKK,TT,DKKMMS,GIR,GutK,Sen,MMV,Kapustin,GKPS,KStro,
DRSVW,KMS2,DD,Alex,GKap,Mukhi,BH}. 
We hence only focus on the case $\hat{c}=5$ ($\cQ=2$) which 
describe a two dimensional target space. 
Generalization to other values of $\hat{c}$ is straightforward.

The type 0 string vacua are defined by the diagonal GSO projection.
Since they are non-supersymmetric, we need not impose the 
integrality of $U(1)$-charge. We may compactify along 
the $Y$-direction with discretized  radius $R=M\cQ\equiv 2M$ ($M\in
\bz_{>0}$, in the $\al'=2$ unit) by imposing the locality of 
the Liouville potential terms $S_+$, $S_-$.
The minimal radius $R=2$ ($M=1$) corresponds to the integral $U(1)$-charges
and fits to  the type II
string vacua. More general radius $R=2M$
corresponds to $N=M$, $K=2M$ and yields fractional $U(1)$-charges 
$Q=n/M\,,n\in {\bf Z}$. Since we have no transverse degrees of freedom, 
the $Y$-direction is naturally identified with the Euclidean 
time. In this sense the system we are considering may be regarded 
as a thermal model with discretized temperatures
$T=1/(4\pi M)$,
analogous to the thermal S-brane models \cite{MSY} 
(see also, {\em e.g.\/} \cite{GS1,Sen-RT,Strominger,LLM}).

The type 0 GSO projections are given by 
\begin{itemize}
\item {\bf type 0A}~ : ~ $J_0+\tJ_0 \in 2\bz$~,
 \item {\bf type 0B}~ : ~ $J_0-\tJ_0 \in 2\bz$~. 
\end{itemize}
We now concentrate on the A and B-type boundary states in the type 0B theory.
The B and A-type branes in the type 0A theory can be studied 
in a parallel way.

~

\noindent
{\bf 1. A-branes} ~:~

The A-branes in the type 0B theory are quite easy. 
The class 1 A-branes should be the $D$-instantons and 
the class 2, 3 A-branes correspond to the $D0$-instanton
(Dirichlet along the $Y$-direction, and Neumann along $\phi$-direction).
All we have to do in the calculation of cylinder amplitudes
is to set $N=M$, $K=2M$ rather than $N=1$, $K=2$ in our analyses. 
We denote the Cardy states for the A-branes as
\begin{eqnarray}
\ketA{B;*,\pm}= \ketA{B;*}^{(\sNS)}\pm \ketA{B;*}^{(\sR)}
\end{eqnarray}
For example, for the class 1 branes we obtain
\begin{eqnarray}
&&\braA{B;r,+}e^{-\pi T H^{(c)}} \ketA{B;r',+} 
= \chi_{\msc{\bf G},\,N=M,K=2M}^{(\sNS)}(r'-r;it,0) 
- \chi_{\msc{\bf G},\,N=M,K=2M}^{(\stNS)}(r'-r;it,0)~, \nn
&& \hspace{10cm} (r,r' \in \bz_{M})~.
\end{eqnarray}
The $\tNS$-contribution in the open string channel originates from
the RR-part of the Cardy state.
The oscillator part $\theta_3/\eta^3$ in the extended characters 
is canceled by the 
contributions from superconformal ghosts.
It is obvious that $r$ expresses the position of $D$-instanton 
along the $S^1$ of $Y$-direction.

Note also that 
\begin{eqnarray}
\braA{B;r,-}e^{-\pi T H^{(c)}} \ketA{B;r',+} 
= \chi_{\msc{\bf G}, \, N=M,K=2M}^{(\sNS)}(r'-r;it,0) 
+ \chi_{\msc{\bf G},\, N=M,K=2M}^{(\stNS)}(r'-r;it,0)~,
\end{eqnarray}
which clearly includes the tachyon mode when $r=r'$.
$\ketA{B;r,-}$ is of course identified as the $\bar{D}$ instanton.
$\ketA{B;*,-}$ for the class 2 and 3 branes are similarly interpreted 
as $\bar{D}$1-branes.

The calculations for the class 2 and 3 branes are similarly carried out,
and in all the cases of $D$-$D$ cylinder amplitudes,
we find only the GSO projected\footnote{Here the ``GSO projected'' precisely
   means that the open string $U(1)$-charge satisfies
$$
J_0\in Q'-Q + 2\bz+1~,
$$
where $Q$, $Q'$ are the $U(1)$-charges of the representations 
labeling the Cardy states.} 
   NS open strings and
no R open strings. The open string vacuum states
have conformal weights  $h \geq 1/2$,
implying that these $D$-branes are stable.
However, the $D$-$\bar{D}$ systems could be unstable as in the class 1
cases. We specially need a careful analysis to examine the stability 
of the class 3 $D$-$\bar{D}$ systems for general $M$.

Under the decompactification limit 
$M\,\rightarrow\,\infty$, we have the continuous spectrum of 
$U(1)$-charges and the extended characters are reduced to the 
irreducible ones \eqn{massive character}, 
\eqn{massless character 1} and \eqn{massless character 2},
as we already noted. For the modular bootstrap
we should use the formulas \eqn{Liouville massive S cont},
\eqn{Liouville massless S 1 cont} and \eqn{Liouville massless S 2 cont}. 
The analysis is almost parallel and the  Cardy states 
are defined associated to the irreducible representations 
$\ch{(*)}{}(h,j_0,n;\tau,z)$, $\chm{(*)}(\la,n;\tau,z)$,
$\chg{(*)}(n;\tau,z)$ of section 2.
There is one difference from the discrete case: 
we can consistently 
define the class 3 states associated with each of 
the massless matter representations $\chm{(*)}(\la,n;\tau,z)$. 
This is because the S-transformation formula
\eqn{Liouville massless S 1 cont} does not contain the ``boundary
term'' contrary to \eqn{Liouville massless S 1}.
The cylinder amplitudes are similarly evaluated. For example, 
we have 
\begin{eqnarray}
\braA{B;n,+}e^{-\pi T H^{(c)}} \ketA{B;n',+} 
= \chg{(\sNS)}(n'-n;it,0) - \chg{(\stNS)}(n'-n;it,0)~,~~~
  (n,n' \in \bz)~,
\end{eqnarray}
for the class 1 branes.

~

\noindent
{\bf 2. B-branes} ~:~

The B-branes in the type 0B theory is more non-trivial.
We first consider the theory of the minimal radius $R=2$ ($M=1$).
In this case $U(1)$-charge for the NSNS (RR) Ishibashi states are integral
(half odd integral). On the other hand, the type 0B GSO projection forces 
all the B-type Ishibashi states to have integral charges. Thus the GSO projection
eliminates all the RR-sector of the theory. 
We thus obtain the general Cardy states for the B-branes
$\ket{B;*}_{\msc{B},\,R=2}$ constructed 
from the B-type Ishibashi states only of the NSNS sector. 
The B-branes so constructed possess  no RR-charges and are
regarded as the generalizations of the non-BPS $D$-branes
in the flat space-time (see \cite{Sen-review}, for instance).
The class 1 B-branes should be $D0$-branes and 
the class 2, 3 B-branes should be $D1$-branes.

The cylinder amplitudes only contain the NS open strings without 
the GSO projection, as in the non-BPS $D$-branes in the flat background.
We have the unique class 1 state $\ket{B;r=0}_{\msc{B},\,R=2}$,
and its overlap is evaluated as 
\begin{eqnarray}
{}_{R=2,\,\msc{B}}\bra{B;0}e^{-\pi T H^{(c)}} \ket{B;0}_{\msc{B},\,R=2} 
= \chi_{\msc{\bf G},\,N=1,K=2}^{(\sNS)}(r=0;it,0) ~.
\end{eqnarray}
This is tachyonic\footnote
   {Since we are now working in the Euclidean theory with no transverse
    degrees of freedom, we do not have the mass shell condition 
    in the usual sense. 
    We look at the IR behavior of the open string amplitudes 
  (after making the oscillator part canceled out with the ghost sector)
   $$
    \chi(it) \sim e^{-2\pi t m^2}~,~~(t\,\rightarrow\,\infty)~,
   $$
    and regard it as massive if $m^2>0$, massless if $m^2=0$, and 
    tachyonic if $m^2<0$.} 
as is expected.
On the other hand, all the class 2, 3 branes contain  
no tachyonic open string states
despite the lack of open string GSO projection
(under the assumption \eqn{bound class 2} for the class 2-branes) 
because of the mass gap $\cQ^2/8=1/2$. 
See Appendix D for the class 3 branes.

The Cardy states for general radius $R=2M$ are obtained 
by making the ``thermal projection operator''  $P_M$ act on 
$\ket{B;*}_{\msc{B},\,R=2}$, as in the construction of the S-brane boundary 
states at finite temperature \cite{MSY};
\begin{eqnarray}
\ket{B;*}_{\msc{B},\,R=2M}= \cC P_M \ket{B;*}_{\msc{B},\,R=2}~.
\end{eqnarray}
$\cC$ is a normalization constant to be fixed by the Cardy condition 
and $P_M$ is the projection operator to the sectors generated 
by the vacua with the following spectrum of $U(1)$-charges 
\begin{eqnarray}
J_0= \frac{n}{M}+2Mw~,~~~\tJ_0=\frac{n}{M}-2Mw~,~~~ (n,w \in \bz)~.
\end{eqnarray}

We can calculate their overlaps in a manner similar to \cite{Sugawara}:
thanks to the B-type boundary condition (Neumann along the $Y$-direction),
one may replace the projection $P_M$ with a simpler one
\begin{eqnarray}
\frac{1}{M}\sum_{s\in \bsz_M}\, e^{i\pi \frac{s}{M}(J_0-\tJ_0)}~,
\label{projection PM}
\end{eqnarray} 
in the amplitudes. We obtain for the class 1 brane, for example, 
\begin{eqnarray}
&& {}_{R=2M,\,\msc{B}}\bra{B;0}e^{-\pi T H^{(c)}} \ket{B;0}_{\msc{B},\,R=2M} 
={|\cC|^2\over M} \sum_{s\in \bsz_{M}}\,
e^{-2\pi t\cdot \frac{5}{2}\left(\frac{s}{M}\right)^2}
\chi_{\msc{\bf G},\,N=1,K=2}^{(\sNS)}
\left(r=0;it,-it\left(s/M\right)\right) ~. \nn
&&
\label{type 0 B-brane}\end{eqnarray}
Thus we should choose $\cC=\sqrt{M}$. 
One can see that the character function 
in the open string channel is generated by 
a fractional spectral flow with $\eta=s/M$ in \eqn{spectral flow}. 
This is because the projection \eqn{projection PM} works as 
twistings  along the spatial direction in the open string 
channel under the modular transformation.  
We note that, even if we have 
closed string states with only integral $U(1)$-charges, 
fractional $U(1)$-charges 
may appear in the open string 
spectrum. We again find an open string tachyon in (\ref{type 0 B-brane}).

In the limit $M\, \rightarrow \,\infty$ we instead obtain
\begin{eqnarray}
{}_{R=\infty,\,\msc{B}}\bra{B;0}e^{-\pi T H^{(c)}} 
\ket{B;0}_{\msc{B},\,R=\infty} 
&=&  |\cC|^2\, \int_0^1 dx\,
e^{-2\pi t \cdot \frac{5}{2}x^2}
\chi^{(\sNS)}_{\msc{\bf G},\, N=1,K=2}\left(r=0 ;it,-it x \right) \nn
&\equiv & |\cC|^2\, \int_{-\infty}^{\infty} dx\,
e^{-2\pi t \cdot \frac{5}{2}x^2}
\chg{(\sNS)}\left(it,-it x \right) ~,
\end{eqnarray}
and hence we must set $\cC=1$ in this case.
The spectral flow parameter becomes continuous  
in this decompactification limit, 
and this behavior is consistent with the fact that 
the $U(1)$-direction (Euclidean time)
is the Neumann boundary condition for the B-branes.

The calculations for the class 2 and 3 branes are similar.
All we have to do is to perform a sum of the results
for $R=2$ over the fractional spectral flows $U_{s/M}$. 
The stabilities of these branes again follow from the mass gap.

In conclusion, the class 1 B-branes are identified as the unstable 
$D0$-branes,
which presumably possess a matrix model description 
and the class 2 and 3 B-branes are the $D1$-branes, 
which are stable due to the mass gap.
~


We would like to present several remarks.

\noindent
{\bf 1.}~
Let us briefly discuss the type IIB case. (Similar results in
the IIA case are obtained by exchanging the roles of A and B-branes.)
This time only the minimal radius $R=2$ is allowed because of 
the locality of space-time supercharges.
The A-branes are of course the BPS branes we treated in the 
previous sections, while the B-branes are analogues of the non-BPS
branes in the flat background. 
The boundary states for the IIB B-branes are the same as
the 0B B-branes given above, but we must insert
the GSO projection $\dsp \frac{1+(-1)^F}{2}$ in the overlap 
amplitudes. The net effect amounts to providing the R-sector
in the open string channel. For example, we 
obtain for the unique class 1 state \footnote
   {Precisely speaking, when calculating thermal amplitudes, 
    we need to take care of the boundary condition for space-time 
    fermions along the thermal circle ($Y$-direction) 
    (see, for instance, \cite{AW}). We shall here omit it 
    to avoid unessential complexity.},
\begin{eqnarray}
&& 2\times {}_{R=2,\,\msc{B}}\bra{B;0}e^{-\pi T H^{(c)}} 
\frac{1+(-1)^F}{2}
\ket{B;0}_{\msc{B},\,R=2} \nn
&& \hspace{1in}
= 
\chi_{\msc{\bf G},\,N=1,K=2}^{(\sNS)}
\left(r=0;it,0\right) 
-\chi_{\msc{\bf G},\,N=1,K=2}^{(\sR)}
\left(r=1;it,0\right)~, 
\end{eqnarray}
where we included the factor $(\sqrt{2})^2\equiv 2$ due to 
the tension of non-BPS $D$-branes.
The amplitude includes the tachyon mode in the NS sector
and the massless mode in the R sector. 
The class 1 B-branes are identified as the non-BPS $D0$-branes
in the IIB vacuum.  Based on the above analysis 
on the overlaps, it seems plausible that these unstable branes possess
a supermatrix model description.
It has been recently proposed in \cite{MMV} that the IIB string vacuum of
the $\cN=2$ Liouville theory should be dual to the supermatrix 
quantum mechanics introduced by Marinari and Parisi \cite{MP}\footnote
     {However, our boundary state of the class 1 B-brane, 
      which we believe to describe the non-BPS $D0$-brane, 
      differs from  the one proposed in \cite{MMV}.
       The latter includes the Ishibashi states with continuous 
       $U(1)$-charge, while the class 1 B-brane only includes
       the ones with discrete $U(1)$-charges. Moreover, they 
       have the different boundary wave functions.}.

Note also that such type II vacua have  
no bose-fermi cancellation in space-time,
although the space-time supercharges are still well-defined 
as local operators. This is because the two dimensional 
Poincar\'{e} invariance is broken due to the Liouville potentials,
as is pointed out in \cite{GKPS}. 
Interestingly, the space-time supercharges possess  nilpotency;
$Q^2=0$ and are interpreted as some BRST charges,
suggesting a topological symmetry  behind the system \cite{DiFK}. 

~


\noindent
{\bf 2.}~ 
The 2d type 0 model considered here has 
the same field content as that of the $c=3/2$ $\cN=1$
superconformal matter coupled to the $\cN=1$ Liouville theory,
studied recently in detail in the context of dual matrix model
\cite{TT,DKKMMS,Kapustin,DRSVW,GKap,BH}. However, one should remember that
the $\cN=2$ Liouville potential $\mu S_++ \bar{\mu}S_-$ is 
the ``sine-Liouville type''\cite{KKK}. It is analogous to the 
inhomogeneous brane decay/creation models considered in  {\em e.g.\/}
\cite{LNT}.
However, an important difference is that our system is already
time dependent {\em in the bulk\/}, because the $\cN=2$ Liouville 
potential breaks the translational invariance $Y\,\rightarrow\, Y+a$.

~

\noindent
{\bf 3.}~
The $\cN=2$ Liouville theory with the minimal radius $R=\cQ(= 2)$ 
is known to be T-dual to the $SL(2;\br)/U(1)$ Kazama-Suzuki model
\cite{GK,HK1} (see also \cite{KKK,FZZ2}). 
In a recent paper \cite{GKPS}, it has been conjectured that 
the 2d type 0 model considered here with $R=2$  
should be dual to the $\bz_2$-symmetric version of 
the KKK matrix model at the self-dual radius \cite{KKK}.
This conjecture is inspired by the proposal
for the $\cN=1$ case given in \cite{TT,DKKMMS},
and based on the similarity of $\cN=2$ Liouville to the sine-Liouville
theory.
It may be interesting 
to study the models with $R=2M$ and to explore 
the relations to the ``symmetric KKK matrix models'' 
with general radii.


~

\section{Summary and Discussions}

~

In this paper we have investigated the boundary states 
in the $\cN=2$ Liouville theory with arbitrary
rational central charge $\hat{c}=1+2K/N$
based on the modular bootstrap approach.
The key idea is to take a sum over spectral flows in order to 
yield discrete spectra of $U(1)$-charges
both in the closed and open string channels. This property
is necessary to construct the BPS $D$-branes in superstring vacua. 
Our modular bootstrap approach is able to determine only 
the absolute values of boundary wave functions and 
we have fixed their phase factors by making use the proposal 
of reflection amplitudes given in \cite{BF,AKRS}.
These boundary wave functions show several nice features
for their identifications as localized/extended $D$-branes.
In any case, we emphasize that all the informations necessary 
for the derivation of cylinder amplitudes can be obtained by our method,
and we can classify consistent $D$-branes.
Application of the method 
of the bootstrap for disk correlation functions \cite{FZZ,ZZ}
for ${\cal N}=2$ Liouville theory will be a difficult but
an important task and will be complementary to our analysis.

We have clarified the analogues of the FZZT-branes (extended along the 
Liouville direction) and the ZZ-branes (localized near the strong 
coupling region $\phi \, \sim \,+\infty$). Among others, 
the analogues of ZZ-branes play important roles in describing
the BPS $D$-branes wrapped around the vanishing cycles in the 
Calabi-Yau singularity, and we have shown that the correct 
intersection numbers are reproduced by the boundary state calculus.

We have also studied briefly the non-BPS $D$-branes in the 
2d type 0 (or type II) string vacuum constructed from 
the $\cN=2$ Liouville with $\hat{c}=5$. Aspects of unstable branes 
are quite similar to those of the $\cN=1$ model of 2d type 0
string \cite{TT,DKKMMS}. This is not surprising since they share the 
same field content as world-sheet theories.
However, the $\cN=2$ 2d type 0 model may be more challenging, 
since the {\em closed\/} string background is already time-dependent 
due to the $\cN=2$ Liouville potential. 
The ``time-like $\cN=2$ Liouville theory''
defined by the Wick rotation of $Y$-direction
would be an interesting problem 
which may shed new light on the study of time-dependent string dynamics.
It may be  worth pointing out that 
the time-like $\cN=2$ Liouville are free from the subtlety
in the analytic continuation of coupling constant
in contrast to the bosonic Liouville case (see \cite{GS2,ST,Schomerus}) where the Liouville
field $\phi$ is regarded as the time direction.


As we have already mentioned, the $\cN=2$ Liouville theory with 
the minimal radius $R=\cQ$ 
is known to be T-dual to the $SL(2;\br)/U(1)$ Kazama-Suzuki model.
It is thus an important problem to compare 
our result with those of the $SL(2;\br)/U(1)$ model.
Recently, in \cite{RibS}, 
the $D$-branes in the bosonic $SL(2;\br)/U(1)$ model
(2d black-hole model) have been studied systematically based on 
the solutions of boundary bootstrap equations in the 
$SL(2;\br)$-WZW model given in \cite{PST} (see also \cite{GKSch,LOP}).
Especially, the ZZ-brane analogues localized at the tip of cigar
in the 2d black-hole geometry have been constructed. 
It is natural to expect that supersymmetric version of these states 
are identified with our class 1 branes in the $\cN=2$ Liouville
theory by T-duality. 
Establishing the precise correspondence 
between $SL(2;\br)/U(1)$ and $\cN=2$ Liouville
with respect to stable/unstable D-branes 
is an important open problem.

In this article we have introduced an Ansatz for the allowed set of 
Ishibashi states based on the considerations of their modular 
properties. It will be better if one could justify this Ansatz 
directly from the analysis of the closed string spectrum of the theory.
We hope that we can report progress on this issue in a future 
publication.

~

\noindent
{\bf Note added :}
After this paper has been submitted to hep-th, 
a new preprint \cite{STT} has appeared
which discusses characters of some affine Lie superalgebra 
and contains results with some overlap with ours. 
We thank A. Taormina for pointing out this reference.

~


\section*{Acknowledgements}
\indent

We are grateful to Y. Nakayama and S. Yamaguchi for valuable discussions.
We also thank the organizers of Summer Institute ``Fuji 2003''
in Shizuoka, Japan for its stimulating atmosphere,
where part of this work was done. 

The research of T. E. and Y. S. is partially  supported by 
Japanese Ministry of Education, 
Culture, Sports, Science and Technology.

\vspace{2cm}

\newpage
\section*{Appendix A ~ Notations}
\setcounter{equation}{0}
\def\theequation{A.\arabic{equation}}

\noindent
{\bf 1. Theta functions}

We here summarize our notations of theta functions.
We set  $q\equiv e^{2\pi i \tau}$ and  $y\equiv e^{2\pi i z}$, 
 \begin{eqnarray}
&& \theta_1(\tau,z) =i\sum_{n=-\infty}^{\infty}(-1)^n q^{(n-1/2)^2/2} y^{n-1/2}
  \equiv 2 \sin(\pi z)q^{1/8}\prod_{m=1}^{\infty}
    (1-q^m)(1-yq^m)(1-y^{-1}q^m), \nn
&&  \theta_2(\tau,z)=\sum_{n=-\infty}^{\infty} q^{(n-1/2)^2/2} y^{n-1/2}
  \equiv 2 \cos(\pi z)q^{1/8}\prod_{m=1}^{\infty}
    (1-q^m)(1+yq^m)(1+y^{-1}q^m)~, \nn
&& \theta_3(\tau,z)=\sum_{n=-\infty}^{\infty} q^{n^2/2} y^{n}
  \equiv \prod_{m=1}^{\infty}
    (1-q^m)(1+yq^{m-1/2})(1+y^{-1}q^{m-1/2})~, \nn
&& \theta_4(\tau,z)=\sum_{n=-\infty}^{\infty}(-1)^n q^{n^2/2} y^{n}
  \equiv \prod_{m=1}^{\infty}
    (1-q^m)(1-yq^{m-1/2})(1-y^{-1}q^{m-1/2}) ~,
\end{eqnarray}
 \begin{eqnarray}
 \Th{m}{k}(\tau,z)&=&\sum_{n=-\infty}^{\infty}
 q^{k(n+\frac{m}{2k})^2}y^{k(n+\frac{m}{2k})} ~,
 \end{eqnarray}
 \begin{equation}
 \eta(\tau)=q^{1/24}\prod_{n=1}^{\infty}(1-q^n)~.
 \end{equation}

~

\noindent
{\bf 2. q-gamma function}

The ``q-gamma function''  $S_b(x)$ is defined as 
\cite{FZZ};
\begin{eqnarray}
\ln S_b(x) = \int_0^{\infty}\frac{dt}{t}\left\{
\frac{\sinh\left(\left(b+\frac{1}{b}-2x\right)t\right)}
{2\sinh(bt)\sinh\left(t/b\right)} -\frac{b+\frac{1}{b}-2x}{2t}
\right\}~, ~~~ (\mbox{for}~ 0<\Re\,x< b+\frac{1}{b})~,
\label{q-gamma}
\end{eqnarray}
and is analytic continued to other regions 
of complex $x$-plane.
This function has simple zeros at $\dsp x= \frac{m+1}{b}+(n+1)b$ 
and simple poles at $\dsp x= -\frac{m}{b}-nb$ ($m,n\in \bz_{\geq 0}$).

~

\noindent
{\bf 3. Character Formulas for $\cN=2$ Minimal Model}

The easiest way to represent the character formulas of 
the level $k$ $\cN=2$ minimal model $(\hat{c}=k/(k+2))$ is to use its
realization as the coset
$\dsp \frac{SU(2)_k\times U(1)_2}{U(1)_{k+2}}$. We then have 
the following branching relation;
\begin{eqnarray}
&& \chi_{\ell}^{(k)}(\tau,w)\Th{s}{2}(\tau,w-z)
=\sum_{\stackrel{m\in \bsz_{2(k+2)}}{\ell+m+s\in 2\bsz}} \chi_m^{\ell,s}
(\tau,z)\Th{m}{k+2}(\tau,w-2z/(k+2))~, \nn
&& \chi^{\ell,s}_m(\tau,z) \equiv  0~, ~~~ \mbox{for $\ell+m+s \in 2\bz+1$}~,
\label{branching}
\end{eqnarray}
where $\chi_{\ell}^{(k)}(\tau,z)$ is the spin $\ell/2$ character of 
$SU(2)_k$;
\begin{equation}
\chi^{(k)}_{\ell}(\tau, z) 
=\frac{\Th{\ell+1}{k+2}(\tau,z)-\Th{-\ell-1}{k+2}(\tau,z)}
                        {\Th{1}{2}(\tau,z)-\Th{-1}{2}(\tau,z)}~.
\label{SU(2) character}
\end{equation}
Then, the desired character formulas are written as 
\begin{eqnarray}
&& \ch{(\sNS)}{\ell,m}(\tau,z) = \chi^{\ell,0}_m(\tau,z)
+\chi^{\ell,2}_m(\tau,z)~, \\
&& \ch{(\stNS)}{\ell,m}(\tau,z) = \chi^{\ell,0}_m(\tau,z)
-\chi^{\ell,2}_m(\tau,z)\equiv 
e^{-i\pi\frac{m}{k+2}}\ch{(\sNS)}{\ell,m}\left(\tau,z+\frac{1}{2}\right)~, \\
&& \ch{(\sR)}{\ell,m}(\tau,z) = \chi^{\ell,1}_m(\tau,z)
+\chi^{\ell,3}_m(\tau,z) \equiv 
q^{\frac{k}{8(k+2)}}y^{\frac{k}{2(k+2)}}
\ch{(\sNS)}{\ell,m+1}\left(\tau,z+\frac{\tau}{2}\right)~, \\
&& \ch{(\stR)}{\ell,m}(\tau,z) = \chi^{\ell,1}_m(\tau,z)
-\chi^{\ell,3}_m(\tau,z) \equiv
- e^{-i\pi\frac{m+1}{k+2}}q^{\frac{k}{8(k+2)}}y^{\frac{k}{2(k+2)}}
\ch{(\sNS)}{\ell,m+1}\left(\tau,z+\frac{1}{2}+\frac{\tau}{2}\right)~. \nn
&&
\end{eqnarray}
By definition, we may restrict to $\ell+m \in 2\bz$ for the  $\NS$ and
$\tNS$ sectors, and to $\ell+m \in 2\bz+1$ for the $\R$ and $\tR$
sectors. The modular transformation coefficients \eqn{minimal S}
can be directly read off from these formulas.

~

\section*{Appendix B ~ A Useful Formula for Modular Calculus}
\setcounter{equation}{0}
\def\theequation{B.\arabic{equation}}
\indent

We here present a useful formula 
relevant for the calculation of S-transformation of 
the massless characters, which has been proved  in \cite{Miki}.
Define the following function
\begin{eqnarray}
 I(k,a,b;\tau,z) \equiv \sum_{r\in \bsz+\frac{1}{2}}\, 
e^{2\pi i a r} \,
\frac{(yq^r)^b}{1+yq^r} \, y^{kr}q^{\frac{k}{2}r^2}~, ~~~
  (a,b\in\br,~ k>0)~,
\label{Ikab}
\end{eqnarray}
then we can prove the following identity\footnote
    {If the integrand in the R.H.S of \eqn{Miki formula} has a pole
     on the real axis, we must regard  the integral as the
     principal value.}
\begin{eqnarray}
&&\frac{i}{\tau} e^{-ik\frac{\pi z^2}{\tau}}\, 
I(k,a,b;-\frac{1}{\tau},\frac{z}{\tau})
=\sum_{r\in\bsz+a}\,e^{i\pi (r-a)}y^r q^{\frac{r^2}{2k}}\,
\frac{1}{\sqrt{k}} \int_{-\infty}^{\infty}dp\,
\frac{e^{-2\pi b \left(\frac{p}{\sqrt{k}}+i\frac{r}{k}\right)}}
{1+e^{-2\pi \left(\frac{p}{\sqrt{k}}+i\frac{r}{k}\right)}}\, 
q^{\frac{p^2}{2}}  \nn
&& \hspace{3cm} + i \sum_{\stackrel{s\in\bsz+\frac{1}{2}}
    {\delta(a,s)\neq 0}}\, e^{i\pi(\delta(a,s)-a)} 
e^{2\pi i \left(\frac{k}{2}-b\right)s} \,
\frac{(yq^s)^{\delta(a,s)}}{1+yq^s}\, y^{ks}q^{\frac{k}{2}s^2} \nn
&& \hspace{3cm}
+\frac{i}{2}\sum_{\stackrel{s\in \bsz+\frac{1}{2}}{\delta(a,s)=0}}
\, e^{-i\pi a}
e^{2\pi i \left(\frac{k}{2}-b\right)s} \,
\frac{1-yq^s}{1+yq^s}\,  y^{ks}q^{\frac{k}{2}s^2}~,
\label{Miki formula}
\end{eqnarray}
where $\delta(a,s)$ is a real number uniquely determined by 
the conditions;
\begin{eqnarray}
&& \delta(a,s)\equiv a-ks ~(\mod\, \bz) ~,\nn
&& 0\leq \delta(a,s)< 1~. 
\end{eqnarray}
(We have  slightly modified  the original notations 
in \cite{Miki} for our convenience.)


~

\section*{Appendix C ~ Modular Transformation Formulas of \\ Characters
of the Extended Chiral Algebras}
\setcounter{equation}{0}
\def\theequation{C.\arabic{equation}}
\indent

We here summarize the modular transformation formulas of 
characters of the extended chiral algebra defined by adding 
the spectral flow operators to the $\cN=2$ algebra in the cases $\hat{c}=2,3,4,5$.
The most familiar example is $\hat{c}=2$ case. The extended algebra is 
nothing but the $\cN=4$ SCA of level 1 and is relevant for 
the $K3$ compactification \cite{ET,EOTY}. 
Generalizations to $\hat{c}=3$ ($CY_3$ compactification)
and $\hat{c}=4$ ($CY_4$ compactification) are discussed in
\cite{Odake,HS}. 

The desired characters are obtained by summing up the irreducible
characters of $\cN=2$ SCA \eqn{massive character},
\eqn{massless character 1}, \eqn{massless character 2}
over the integral spectral flows.
\begin{eqnarray}
\Ch{(\sNS)}{}(h,Q;\tau,z) &= & \sum_{m\in\bsz}
q^{\frac{\hat{c}}{2}m^2}y^{\hat{c}m}
\ch{(\sNS)}{}(h,Q;\tau,z+m\tau) \nn
&=& q^{h-\frac{Q^2}{2(\hat{c}-1)}-\frac{\hat{c}-1}{8}}
\,\Theta_{Q,\frac{\hat{c}-1}{2}}(\tau,2z)\, 
\frac{\theta_3(\tau,z)}{\eta(\tau)^3}~,
\label{extended character 1}\\
\Chm{(\sNS)}(Q(\neq 0);\tau,z) &= & \sum_{m\in\bsz}\,
q^{\frac{\hat{c}}{2}m^2}y^{\hat{c}m}
\chm{(\sNS)}(Q;\tau,z+m\tau) \nn
&=& q^{-\frac{\hat{c}-1}{8}} \sum_{m\in\bsz}\,
\frac{q^{\frac{\hat{c}-1}{2}m^2+|Q|m+\frac{|Q|}{2}} 
y^{\msc{sgn(Q)}(|Q|+(\hat{c}-1)m)}}{1+y^{\msc{sgn}(Q)}q^{m+1/2}}\, 
\frac{\theta_3(\tau,z)}{\eta(\tau)^3}~, 
\label{extended character 2}\\
\Chg{(\sNS)}(\tau,z) &= & \sum_{m\in\bsz}\,
q^{\frac{\hat{c}}{2}m^2}y^{\hat{c}m}
\chg{(\sNS)}(\tau,z+m\tau) \nn
&=& q^{-\frac{\hat{c}-1}{8}} \sum_{m\in\bsz}\,
\frac{(1-q)q^{\frac{\hat{c}-1}{2}m^2+m-\frac{1}{2}}y^{(\hat{c}-1)m+1}}
{(1+yq^{m+1/2})(1+yq^{m-1/2})}\,
\frac{\theta_3(\tau,z)}{\eta(\tau)^3}~. 
\label{extended character 3}
\end{eqnarray}
Note that, in the cases of $\hat{c}=3,5$ these characters themselves are 
identified as the extended characters $\chi_*(*;\tau,z)$ 
defined in section 2, while in the other cases 
$\hat{c}=2,4$ one takes the sum
$\chi_*(r=0,*;\tau,z)+\chi_*(r=-1,*;\tau,z)$ to define the characters of the extended algebra. Furthermore, the sectors with odd quantum numbers $j$, $s$
are eliminated by the locality with the integral spectral flow 
generators $U_{\pm 1}$, rather than $U_{\pm 2}$.

The extended characters for other spin structures are defined 
by the 1/2-spectral flow\footnote
    {We  shall here adopt the convention 
     such that  the quantum numbers appearing in  
     the Ramond characters are the same as the corresponding
      NS ones. Namely, $(h,Q)$ in the Ramond characters 
      are {\em not\/} equal the conformal weights and $U(1)$-charges 
      of the Ramond vacuum states. Although this convention may be 
      somewhat confusing, it is convenient to write down the modular 
      transformation formulas for general spin structures.};
\begin{eqnarray}
\Ch{(\stNS)}{*}(*;\tau,z) &\equiv& \Ch{(\sNS)}{*}(*;\tau,z+\frac{1}{2})~, \nn
\Ch{(\sR)}{*}(*;\tau,z) &\equiv&q^{\hat{c}/8}y^{\hat{c}/2} 
\Ch{(\sNS)}{*}(*;\tau,z+\frac{\tau}{2}) ~, \nn
\Ch{(\stR)}{*}(*;\tau,z) &\equiv&q^{\hat{c}/8}y^{\hat{c}/2} 
\Ch{(\sNS)}{*}(*;\tau,z+\frac{\tau}{2}+\frac{1}{2})~.
\end{eqnarray}
We here used the abbreviated notations which should be understood 
in the meaning as in \eqn{extended character 1}, 
\eqn{extended character 2} or \eqn{extended character 3}.

First of all, it is quite easy to evaluate the $T$-transformation. 
We obtain 
\begin{eqnarray}
\Ch{(\sigma)}{}(h,Q;\tau+1,z)&=&e^{2\pi i \gamma(h,\sigma)}\,
\Ch{(T\cdot \sigma)}{}(h,Q;\tau,z) ~, \nn
\Chm{(\sigma)}(Q;\tau+1,z)&=&e^{2\pi i \gamma(|Q|/2,\sigma)}\,
\Chm{(T\cdot \sigma)}(Q;\tau,z)~, \nn
\Chg{(\sigma)}(\tau+1,z)&=&e^{2\pi i \gamma(0,\sigma)}\,
\Chg{(T\cdot \sigma)}(\tau,z)~,
\label{T transformation}
\end{eqnarray}
where we set 
\begin{eqnarray}
\gamma(h,\sigma) \equiv 
\left\{
\begin{array}{ll}
 h-\frac{\hat{c}}{8}&   ~~(\sigma=\NS,\,\tNS) \\
 h & ~~ (\sigma=\R,\,\tR)
\end{array}
\right.
\end{eqnarray}
and $T\cdot \NS=\tNS$, $T\cdot \tNS = \NS$, $T\cdot \R=\R$, $T\cdot \tR=\tR$.

Let us next consider the S-transformation.
For the massive representations, we have 
the following continuous spectra. The unitarity condition 
is $h>|Q|/2$ in all the cases.
\begin{itemize}
 \item $\hat{c}=2$ : ~ We have 
                   only one continuous series generated by the highest-weights 
                      state $(h,Q=0)$.
 \item $\hat{c}=3$ : ~ We  have two continuous series $(h,Q=0)$, 
                       $(h,Q=\pm 1)$.
                       (the vacua are doubly degenerated in the second
                        case).
 \item $\hat{c}=4$ : ~ We have  three continuous series $(h,Q=0)$, $(h,Q=+1)$
                  and $(h,Q=-1)$.
 \item $\hat{c}=5$ :~ We have four continuous series $(h,Q=0)$, $(h,Q=+1)$, 
                      $(h,Q=-1)$ and $(h,Q=\pm 2)$ (the vacua are doubly
                        degenerated in the last case).
\end{itemize}

Since the massive character is written as in \eqn{extended character 1},
the S-transformation formula is easily follows from that of the 
theta function; 
\begin{eqnarray}
&& \Ch{(\sigma)}{}\left(h=\frac{p^2}{2}+\frac{Q^2}{2(\hat{c}-1)}
+\frac{\hat{c}-1}{8}, Q;-\frac{1}{\tau}, \frac{z}{\tau}\right) 
=\kappa(\sigma)e^{i\pi \frac{\hat{c} z^2}{\tau}}\, \frac{2}{\sqrt{\hat{c}-1}}
\int_{0}^{\infty}dp'\, \cos(2\pi pp')\,  \nn
&& \hspace{1cm} \times
\sum_{Q' \in \bsz_{\hat{c}-1}}\, e^{-2\pi i \frac{Q Q'}{\hat{c}-1}}
\Ch{(S\cdot\sigma)}{}\left(h=\frac{p^{'2}}{2}+\frac{Q^{'2}}{2(\hat{c}-1)}
+\frac{\hat{c}-1}{8}, Q';\tau,z\right)~, 
\label{extended massive S}
\end{eqnarray}
where we set 
\begin{eqnarray}
\kappa(\sigma)=\left\{
\begin{array}{ll}
 1&~~\sigma=\NS,\, \tNS,\, \R \\
 e^{-i\pi \frac{\hat{c}}{2}}&~~\sigma=\tR
\end{array}
\right.~,
\end{eqnarray}
and $S\cdot \NS=\NS$, $S\cdot \tNS=\R$, $S\cdot \R=\tNS$,
$S\cdot \tR=\tR$.

The formulas for the massless characters are much more non-trivial. 
We can evaluate them by using 
the formula \eqn{Miki formula}. They are summarized as follows;

\noindent
{\bf 1. ~$\hat{c}=2$} : 

We have two massless representations
$Q=0$ (graviton representation) and $Q=\pm 1$ (doubly degenerated vacua).
The modular transformation formulas are written as  \cite{ET};
\begin{eqnarray}
&& \Chg{(\sigma)}\left(-\frac{1}{\tau}, \frac{z}{\tau}\right)
= \kappa(\sigma)e^{i\pi \frac{2 z^2}{\tau}}\,\left\{
2\Chm{(S\cdot\sigma)}(|Q|=1;\tau,z)   \right. \nn
&&\hspace{1cm}
\left. + 2 \int_{0}^{\infty}dp'\, \sinh(\pi p')\tanh(\pi p')\,
\Ch{(S\cdot \sigma)}{}\left(h=\frac{p^{'2}}{2}+\frac{1}{8}, Q=0;\tau,z\right)
\right\} ~,\\
&& \Chm{(\sigma)}\left(|Q|=1;-\frac{1}{\tau}, \frac{z}{\tau}\right)
= \kappa(\sigma)e^{i\pi \frac{2 z^2}{\tau}}\,\left\{
-\Chm{(S\cdot\sigma)}(|Q|=1;\tau,z)   \right. \nn
&&\hspace{1cm}
\left. + 
\int_{0}^{\infty}dp'\, \frac{1}{\cosh(\pi p')}\,
\Ch{(S\cdot \sigma)}{}\left(h=\frac{p^{'2}}{2}+\frac{1}{8}, Q=0;\tau,z\right)
\right\} ~.
\end{eqnarray}

~

\noindent
{\bf 2. ~ $\hat{c}=3$} :

We have three massless representations $Q=0$, $Q=+1$ and $Q=-1$.
The modular transformation formulas are written as  \cite{Odake};
\begin{eqnarray}
&& \Chg{(\sigma)}\left(-\frac{1}{\tau},
\frac{z}{\tau}\right)
= \kappa(\sigma)e^{i\pi \frac{3 z^2}{\tau}} \sqrt{2} 
\int_{0}^{\infty}dp'\, \sinh\left(\sqrt{2}\pi p'\right) \nn
&& \hspace{2cm} \times
\left\{ 
\tanh\left(\frac{\pi p'}{\sqrt{2}}\right)\,
\Ch{(S\cdot\sigma)}{}\left(h=\frac{p^{'2}}{2}+\frac{1}{4}, Q=0;\tau,z\right)
\right.  \nn
&& \hspace{2cm}
\left. 
+ \coth\left(\frac{\pi p'}{\sqrt{2}}\right)\,
\Ch{(S\cdot\sigma)}{}\left(h=\frac{p^{'2}}{2}+\frac{1}{2}, |Q|=1;\tau,z\right)
\right\}~, \nn
&& \\
&& \Chm{(\sigma)}\left(Q=\pm 1;-\frac{1}{\tau},
\frac{z}{\tau}\right)=\kappa(\sigma)e^{i\pi \frac{3z^2}{\tau}} 
\nn
&&  
\times \left\lb 
\frac{1}{\sqrt{2}}\int_{0}^{\infty}dp'\, 
\left\{
\Ch{(S\cdot\sigma)}{}\left(h=\frac{p^{'2}}{2}+\frac{1}{4},Q=0;\tau,z\right)
-\Ch{(S\cdot\sigma)}{}\left(h=\frac{p^{'2}}{2}+\frac{1}{2},|Q|=1;\tau,z\right)
\right\} \right. \nn
&& \hspace{2cm} \left.
-\frac{i}{2}\left\{
\Chm{(S\cdot\sigma)}(Q=\pm 1;\tau,z)-
\Chm{(S\cdot\sigma)}(Q=\mp 1;\tau,z)
\right\} \right\rb ~.
\end{eqnarray}

~

\noindent
{\bf 3. ~ $\hat{c}=4$} :

We have four massless representations $Q=0$, $Q=+1$, $Q=-1$
and $Q=\pm 2$ (doubly degenerated vacua).
The modular transformation formulas are written as;
\begin{eqnarray}
&& \Chg{(\sigma)}\left(-\frac{1}{\tau}, \frac{z}{\tau}\right)
=e^{i\pi \frac{4 z^2}{\tau}} \left\lb 
2\Chm{(S\cdot\sigma)}(|Q|=2;\tau,z) 
\right. \nn
&& \hspace{1cm} +\frac{2}{\sqrt{3}}\int_{0}^{\infty}dp'\,
\sinh(\sqrt{3}\pi p')\tanh\left(\frac{\pi p'}{\sqrt{3}}\right)\,
\Ch{(S\cdot\sigma)}{}\left(h=\frac{p^{'2}}{2}+\frac{3}{8},Q=0;\tau,z\right)
\nn
&& \hspace{1cm} 
+\frac{1}{\sqrt{3}}\int_{0}^{\infty}dp'\,
\frac{\sinh(\sqrt{3}\pi p')\sinh\left(\frac{2\pi p'}{\sqrt{3}}\right)}
{\left|\cosh\left(\frac{\pi p'}{\sqrt{3}}+\frac{i\pi}{3}\right)\right|^2}
\nn
&& \hspace{1cm}\left. \times
\left\{
\Ch{(S\cdot\sigma)}{}\left(h=\frac{p^{'2}}{2}+\frac{13}{24},Q=1;\tau,z\right)
+\Ch{(S\cdot\sigma)}{}\left(h=\frac{p^{'2}}{2}+\frac{13}{24},Q=-1;\tau,z\right)
\right\} \right\rb ~, \nn
&& \\
&& \Chm{(\sigma)}\left(Q=\pm 1;-\frac{1}{\tau}, \frac{z}{\tau}\right)
=e^{i\pi \frac{4 z^2}{\tau}} \left\lb 
- \Chm{(S\cdot\sigma)}(|Q|=2;\tau,z) 
\right. \nn
&& \hspace{1cm} +\frac{1}{2\sqrt{3}}\int_{0}^{\infty}dp'\, \left\{
\frac{2\cosh\left(\frac{2\pi p'}{\sqrt{3}}\right)}
{\cosh\left(\frac{\pi p'}{\sqrt{3}}\right)}\,
\Ch{(S\cdot\sigma)}{}\left(h=\frac{p^{'2}}{2}+\frac{3}{8},Q=0;\tau,z\right)
\right. \nn
&& \hspace{1cm} 
- \frac{e^{i\frac{\pi}{3}}\cosh(\sqrt{3}\pi p')
-\cosh\left(\frac{\pi p'}{\sqrt{3}}\right)}
{\left|\cosh\left(\frac{\pi p'}{\sqrt{3}}+i\frac{\pi}{3}\right)\right|^2}  \,
\Ch{(S\cdot\sigma)}{}\left(h=\frac{p^{'2}}{2}+\frac{13}{24},
Q=\pm 1;\tau,z\right)  \nn
&& \hspace{1cm} \left. \left.
- \frac{e^{-i\frac{\pi}{3}}\cosh(\sqrt{3}\pi p')
-\cosh\left(\frac{\pi p'}{\sqrt{3}}\right)}
{\left|\cosh\left(\frac{\pi p'}{\sqrt{3}}+i\frac{\pi}{3}\right)\right|^2}  \,
\Ch{(S\cdot\sigma)}{}\left(h=\frac{p^{'2}}{2}+\frac{13}{24},
Q=\mp 1;\tau,z\right)  \right\} \right\rb ~, \\
&& \Chm{(\sigma)}\left(|Q| =2;-\frac{1}{\tau}, \frac{z}{\tau}\right)
=e^{i\pi \frac{4 z^2}{\tau}} \left\lb 
\Chm{(S\cdot\sigma)}(|Q|=2;\tau,z) 
\right. \nn
&& \hspace{1cm} +\frac{1}{2\sqrt{3}}\int_{0}^{\infty}dp' \,
\left\{
\frac{2}{\cosh\left(\frac{\pi p'}{\sqrt{3}}\right)}\,
\Ch{(S\cdot\sigma)}{}\left(h=\frac{p^{'2}}{2}+\frac{3}{8},Q=0;\tau,z\right)
\right. \nn
&& \hspace{1cm} 
-  \frac{\cosh\left(\frac{\pi p'}{\sqrt{3}}\right)}
{\left|\cosh\left(\frac{\pi p'}{\sqrt{3}}+i\frac{\pi}{3}\right)\right|^2}
\left(
\Ch{(S\cdot\sigma)}{}\left(h=\frac{p^{'2}}{2}+\frac{13}{24},Q=1;
\tau,z\right) 
\right. \nn
&& \hspace{4cm} \left. \left. \left.
+\Ch{(S\cdot\sigma)}{}\left(h=\frac{p^{'2}}{2}+\frac{13}{24},Q=-1;\tau,z\right)
\right) \right\} \right\rb ~.
\end{eqnarray}

~

\noindent
{\bf 4.~ $\hat{c}=5$} :

We have five massless representations $Q=0$, $Q=+1$, $Q=-1$, $Q=+2$
and $Q=-2$. 
The S-transformation formulas are 
\begin{eqnarray}
&& \Chg{(\sigma)}\left(-\frac{1}{\tau}, \frac{z}{\tau}\right)
=\kappa(\sigma)e^{i\pi\frac{5 z^2}{\tau}} \, \int_{0}^{\infty}dp'\,
\sinh (2\pi p')\nn
&& \hspace{5mm} \times \left\lb  
\tanh\left(\frac{\pi p'}{2}\right) \Ch{(S\cdot\sigma)}{}
\left(h=\frac{p^{'2}}{2}+\frac{1}{2}, Q=0;\tau,z\right)
\right. \nn 
&& \hspace{5mm} 
+ \tanh(\pi p') \left\{
\Ch{(S\cdot\sigma)}{}
\left(h=\frac{p^{'2}}{2}+\frac{5}{8}, Q=1;\tau,z\right)
+\Ch{(S\cdot\sigma)}{}
\left(h=\frac{p^{'2}}{2}+\frac{5}{8}, Q=-1;\tau,z\right)
\right\} \nn
&& \hspace{5mm}
\left. 
+ \coth\left(\frac{\pi p'}{2}\right) \Ch{(S\cdot\sigma)}{}
\left(h=\frac{p^{'2}}{2}+1, |Q|=2;\tau,z\right) \right\rb~,\\
&& \Chm{(\sigma)}\left(Q=\pm 1;-\frac{1}{\tau}, \frac{z}{\tau}\right)
=\kappa(\sigma)e^{i\pi\frac{5 z^2}{\tau}} \,\left\lb\, 
 \frac{1}{2}\int_{0}^{\infty}dp'\, \right.\nn
&& \hspace{5mm}
\times \left\{ 
\frac{\cosh\left(\frac{3\pi p'}{2}\right)}
{\cosh\left(\frac{\pi p'}{2}\right)} \Ch{(S\cdot \sigma)}{}
\left(\frac{p^{'2}}{2}+\frac{1}{2}, 0;\tau,z\right) 
+\left(1-\frac{i\cosh(2\pi p')}{\cosh(\pi p')}\right)
\Ch{(S\cdot \sigma)}{}
\left(\frac{p^{'2}}{2}+\frac{5}{8}, \pm 1;\tau,z\right)\right. \nn
&&\hspace{5mm} \left.
+\left(1+\frac{i\cosh(2\pi p')}{\cosh(\pi p')}\right)
\Ch{(S\cdot \sigma)}{}
\left(\frac{p^{'2}}{2}+\frac{5}{8}, \mp 1;\tau,z\right)
-\frac{\sinh\left(\frac{3\pi p'}{2}\right)}
{\sinh\left(\frac{\pi p'}{2}\right)} \Ch{(S\cdot \sigma)}{}
\left(\frac{p^{'2}}{2}+\frac{1}{2}, 2;\tau,z\right) 
\right\}~, \nn
&& \hspace{5mm}\left.
-\frac{i}{2}\left\{
\Chm{(S\cdot \sigma)}(Q=\pm 2;\tau,z)
-\Chm{(S\cdot \sigma)}(Q=\mp 2;\tau,z)
\right\}
\right\rb~, \\
&& \Chm{(\sigma)}\left(Q=\pm 2;-\frac{1}{\tau}, \frac{z}{\tau}\right)
=\kappa(\sigma)e^{i\pi \frac{5 z^2}{\tau}} \, \left\lb\,
 \frac{1}{2}\int_{0}^{\infty}dp'\, \right. \nn
&& \hspace{5mm}
\times \left\{
\Ch{(S\cdot \sigma)}{}
\left(\frac{p^{'2}}{2}+\frac{1}{2}, 0;\tau,z\right) 
-\left(1+\frac{i}{\cosh(\pi p')}\right)
\Ch{(S\cdot \sigma)}{}
\left(\frac{p^{'2}}{2}+\frac{5}{8}, \pm 1;\tau,z\right)\right. \nn
&&\hspace{5mm} \left.
-\left(1-\frac{i}{\cosh(\pi p')}\right)
\Ch{(S\cdot \sigma)}{}
\left(\frac{p^{'2}}{2}+\frac{5}{8}, \mp 1;\tau,z\right)
+\Ch{(S\cdot \sigma)}{}
\left(\frac{p^{'2}}{2}+\frac{1}{2}, 2;\tau,z\right) \right\}\nn
&& \hspace{5mm}\left.
+\frac{i}{2}\left\{
\Chm{(S\cdot \sigma)}(Q=\pm 2;\tau,z)
-\Chm{(S\cdot \sigma)}(Q=\mp 2;\tau,z)
\right\}
\right\rb~, 
\end{eqnarray}

~

\section*{Appendix D ~ Analysis on the Class 3 States : Concrete Examples}
\setcounter{equation}{0}
\def\theequation{D.\arabic{equation}}
\indent

Checking the Cardy condition for the overlaps among the class 3 states
is a non-trivial problem.
Here we try to do this task for the special cases $\hat{c}=2,3,4,5$
(values of $(N,K)$ are given by $(2,1),(1,1),(2,3),(1,2)$, respectively).

Since we are interested in the BPS $D$-branes 
in the type II string vacua: \\ $\br^{d-1,1} 
\times \mbox{$\cN=2$ Liouville}$ ($\hat{c}+d/2=5$),
we shall only treat the sectors with integral $U(1)$-charges
in the closed as well as open string channel. 
This restriction amounts to only considering 
the boundary states preserved by the extended chiral algebras,
and thus the relevant conformal blocks should be  expanded by the 
characters of extended chiral algebras $\Ch{(*)}{}(h,Q;\tau,z)$,
$\Chm{(*)}(Q;\tau,z)$ and $\Chg{(*)}(\tau,z)$. 
Their modular transformation formulas are quite useful 
for the calculation of cylinder amplitudes and 
summarized in Appendix C.
For simplicity, we concentrate on the NSNS part of the boundary states 
and omit the insertion of the GSO projection $\dsp \frac{1+(-1)^F}{2}$,
which amount to only considering 
the NS sector in the open string channel.
One can easily obtain the results for other spin structures
by spectral flow. 

~

\noindent
{\bf 1.}~ $\hat{c}=2$

The $\hat{c}=2$ case corresponds to $N=2$, $K=1$ ($\cQ=1$).
The class 1 and class 2 states  are written as 
\begin{eqnarray}
&& \mbox{\bf class 1 states : }~~
 \ket{B;r=0}+\ket{B;r=-1}~, ~~ \nn
&& \mbox{\bf class 2 states : }~~
 \ket{B;p,j=0}+\ket{B;p,j=2}~~(\frac{p^2}{2}> -\frac{1}{32})~.
\label{Cardy states hat c 2}
\end{eqnarray}
We have taken account of the projection to  
integral $U(1)$-charges as in section 4.
It is obvious that $\ket{B;O}\equiv \ket{B;r=0}$.  
We have also imposed the condition \eqn{bound class 2} for
the class 2 states.
Note that $\ket{B;p,j=\pm 1}$ correspond to the 
massive representation with  half-integral $U(1)$-charges
and thus we exclude them.

Since we are imposing the integrality of $U(1)$-charge,
we just have a unique possibility of the class 3 states
$\lb (0,2),\, (-1,2) \rb$ which corresponds to the combination
of the characters as
\begin{eqnarray}
\chim^{}(r=0,s=2;\tau,z)+\chim^{}(r=-1,s=2;\tau,z)
\equiv \Chm{(\sNS)}(|Q|=1;\tau,z)~.
\label{c relation c 2}
\end{eqnarray}  
We thus denote this state as $\ket{B;|Q|=1}$.
It is easy to calculate the overlap with itself, and 
we find that it cannot be written in the form \eqn{Cardy condition}
with any appropriate spectral density. 
We thus conclude that
Cardy states are given by the set \eqn{Cardy states hat c 2}
and no class 3 states are allowed.

~

\noindent
{\bf 2.} $\hat{c}=3$

$\hat{c}=3$ corresponds to $N=K=1$ ($\cQ=\sqrt{2}$).
The class 1 and class 2 states are given as
\begin{eqnarray}
&& \mbox{\bf class 1 states : }~~
 \ket{B;r=0}\equiv \ket{B;O}~, \nn
&& \mbox{\bf class 2 states : }~~
 \ket{B;p,j=0}~~(\frac{p^2}{2}> -\frac{1}{16})~, ~~~
 \ket{B;p,j=1}~~(\frac{p^2}{2}\geq 0)~.
\label{Cardy states hat c 3}
\end{eqnarray}

We only have a unique possibility of the class 3 state;
$\lb (0,1),\,(0,2)\rb$. However, because the following character 
identity holds;
\begin{eqnarray}
\chim^{}(0,1;\tau,z)+\chim^{}(0,2;\tau,z)
= \chi(p=0,j=1;\tau,z) (\equiv \Ch{(\sNS)}{}(h=1/2,|Q|=1;\tau,z))~,
\label{c relation c 3}
\end{eqnarray}
it is a special case of the class 2 states 
\eqn{Cardy states hat c 3}. 
We thus conclude that only \eqn{Cardy states hat c 3} 
are allowed as the Cardy states.

~

\noindent
{\bf 3.} $\hat{c}=4$

$\hat{c}=4$ corresponds to $N=2$, $K=3$ ($\cQ=\sqrt{3}$).
This is the most non-trivial example.
The class 1 and class 2 state are given as
\begin{eqnarray}
&& \mbox{\bf class 1 states : }~~
 \ket{B;r=0}(\equiv \ket{B;O})+\ket{B;r=-1}  \nn
&& \mbox{\bf class 2 states : }~~
 \ket{B;p,2j}+\ket{B;p,2j+6}
~~(j = 0,1,2 ~,~~ 
\frac{p^2}{2} > -\frac{3}{32} ~~ (j = 0)~, ~~~ \nn
&& \hspace{4cm}
\frac{p^2}{2} \geq  -\frac{1}{24} ~~(j= 1,2 ))~. 
\label{Cardy states hat c 4}
\end{eqnarray}
We again have taken account of the projection to the closed string 
sectors with integral $U(1)$-charges.
Note that $\ket{B;p,2j+1}+\ket{B;p,2j+7}$ are 
eliminated by our assumption for the Cardy states.
All the class 2 states above are compatible with unitarity even in
the cases of imaginary $p$. 

We have many possibilities of the class 3 states. 
Fortunately, many of the pairs defining the class 3 states \eqn{class 3}
are eliminated by the assumption of charge integrality,
or reduced to the class 2 states by the character identities similar to
\eqn{c relation c 3}. Only three candidates remain: 
$\lb (0,2),\, (-1,2)\rb$, $\lb (0,4),\, (-1,4)\rb$,
$\lb (0,6),\, (-1,6)\rb$, which respectively correspond
to the massless characters;
\begin{eqnarray}
&& \chim^{}(0,2;\tau,z)+\chim^{}(-1,2;\tau,z) = \Chm{(\sNS)}(Q=1;\tau,z)~,\nn
&& \chim^{}(0,4;\tau,z)+\chim^{}(-1,4;\tau,z) 
= \Chm{(\sNS)}(|Q|=2;\tau,z)~,\nn
&& \chim^{}(0,6;\tau,z)+\chim^{}(-1,6;\tau,z) = \Chm{(\sNS)}(Q=-1;\tau,z)~,
\label{character relation hat c 4 L}
\end{eqnarray}
and hence we denote them as 
$\ket{B;Q=1}$, $\ket{B;Q=-1}$ and $\ket{B;|Q|=2}$.

However, all of them are found not to yield 
positive spectral densities in their overlap amplitudes.
We thus look for the proper Cardy states satisfying \eqn{Cardy
condition} among the linear combinations
\begin{eqnarray}
\ket{B;(m_+,m_-,n)} = m_+\ket{B;Q=1}+m_-\ket{B;Q=-1}+n\ket{B;|Q|=2}~,
~~ m_+,m_-,n \in \bz_{\geq 0}~.
\label{mmn}
\end{eqnarray}
We set
\begin{eqnarray}
&& \bs_+ = \left\{\ket{B;(m_+,m_-,n)}\,|\, -m_+-m_-+n >0\right\}~,~~\nn
&& \bs_- = \left\{\ket{B;(m_+,m_-,n)}\,|\, -m_+-m_-+n <0\right\}~,~
\end{eqnarray} 
where $m_+,m_-,n \in \bz_{\geq 0}$ and we assume at least two of 
integers $m_+,m_-,n$ are non-zero.
Then, the careful analysis on overlaps leads to the results;
\begin{itemize}
 \item The overlaps among any pair of states \eqn{mmn} both belonging
  to the same set $\bs_+$ (or $\bs_-$) satisfy the Cardy condition
  \eqn{Cardy condition} with the positive spectral densities.
 \item Any pairing between the states of $\bs_+$ and $\bs_-$ 
  cannot yield the positive spectral densities.   
\end{itemize} 
For example, we obtain after a lengthy calculations
\begin{eqnarray}
&& e^{\pi\frac{4z^2}{T}}\,\bra{B;(m_+,m_-,0)} e^{-\pi T H^{(c)}}
   e^{i\pi z(J_0+\tJ_0)} \ket{B;(m_+',m_-',0)} \nn
&& \hspace{5mm} = 
\int_0^{\infty}dp\, \sum_{Q=0,\pm 1}\, 
\rho_Q(p|m_+,m_-,m_+',m_-')
\Ch{(\sNS)}{}\left(h=\frac{p^2}{2}+\frac{Q^2+9}{24},Q;it,z'\right)
 \nn
&& \hspace{3cm} + (m_++m_-)(m_+'+m_-')\Chm{(\sNS)}(|Q|=2;it,z')~, \\
&& \rho_0(p|m_+,m_-,m_+',m_-')
= 2(m_+m_+'+m_-m_-')\int_0^{\infty}dp'\, 
\frac{\cos(2\pi p p')\cosh(\pi \sqrt{3}p')}
{\sinh(\pi \sqrt{3}p')\sinh\left(\frac{\pi p'}{\sqrt{3}}\right)} ~, 
\label{rho mm 1}\\
&& \rho_{\pm 1}(p|m_+,m_-,m_+',m_-')
= \int_0^{\infty}dp'\, 
\frac{\cos(2\pi p p')}
{\sinh(\pi \sqrt{3}p')\sinh\left(\frac{\pi p'}{\sqrt{3}}\right)}  \nn
&& \hspace{1cm} \times 
\left\{
\left((m_++m_-)(m_+'+m_-')-2m_{\pm}m_{\mp}'+2m_{\mp}m_{\pm}'\right)
\cosh\left(\frac{\pi p'}{\sqrt{3}}\right)
\right. \nn
&& \hspace{1cm} 
\left. -\left((m_++m_-)(m_+'+m_-')-2m_{\pm}m_{\mp}'\right)
\cosh(\pi \sqrt{3}p') \right\}~,
\label{rho mm 2}
\end{eqnarray}
where the characters of the $\hat{c}=4$ extended algebra are 
given as 
\begin{eqnarray}
&& \Ch{(\sNS)}{}\left(\frac{p^2}{2}+\frac{Q^2+9}{24},Q;\tau,z\right)
= \chi(p,2Q;\tau,z)+\chi(p,2Q+6;\tau,z)~,
\end{eqnarray}
and also by \eqn{character relation hat c 4 L}.
The spectral densities \eqn{rho mm 1} and \eqn{rho mm 2}
are defined with the IR regularization considered before. 
They are positive for arbitrary $m_+,m_-,m_+',m_-'\in \bz_{>0}$,
and rewritten by using the q-gamma function \eqn{q-gamma}.

We here make a comment: 
we have a character identity 
\begin{eqnarray}
&&\Chm{(\sNS)}(Q=\pm 1;\tau,z)+\Chm{(\sNS)}(|Q|=2;\tau,z)
= \Ch{(\sNS)}{} \left(h=1/2, Q=\pm 1;\tau,z\right) \nn
&&\hspace{1cm} \equiv  \chi\left(p=\frac{i}{2\sqrt{3}}, j=\pm 2;\tau,z\right)
+ \chi\left(p=\frac{i}{2\sqrt{3}}, j=\mp 4;\tau,z\right)~.
\end{eqnarray} 
Therefore, any state of the form 
$\ket{B;(m_+,m_-,n)}$, $-m_+-m_-+n=0$
is reduced to a (linear combination of) class 2 state. 
Moreover, we can identify the minimal ``basis'' 
of the class 3 Cardy states that cannot be decomposed into 
other Cardy states as 
\begin{eqnarray}
&& \ket{B;(1,0,n)}~,~~(n \geq 2)~,~~~
\ket{B;(0,1,n)}~,~~(n \geq 2)~,
\end{eqnarray}
for $\bs_+$, and 
\begin{eqnarray}
&& \ket{B;(m_+,1,0)}~,~~(m_+\geq 1)~, ~~~
   \ket{B;(1,m_-,0)}~,~~(m_-\geq 2)~, ~~~\nn
&& \ket{B;(m_+,0,1)}~,~~(m_+\geq 2)~, ~~~
   \ket{B;(0,m_-,1)}~,~~(m_-\geq 2)~, 
\end{eqnarray}
for $\bs_-$.

~

\noindent
{\bf 4.} $\hat{c}=5$

We finally consider the special case $\hat{c}=5$ in which 
we have no transverse degrees of freedom.
This case corresponds to $N=1$, $K=2$ ($\cQ=2$) and 
quite reminiscent of the $\hat{c}=3$ case.

The class 1 and 2 states are written as 
\begin{eqnarray}
&& \mbox{\bf class 1 states : }~~
 \ket{B;r=0}\equiv \ket{B;O}~, \nn
&& \mbox{\bf class 2 states : }~~
 \ket{B;p,j}~~(j \in \bz_4~,~~
\frac{p^2}{2}> -\frac{1}{8} ~~(j = 0,1,-1)~,~~~ 
\frac{p^2}{2}\geq 0 ~~(j=2) ). ~ \nn
&& 
\label{Cardy states hat c 5}
\end{eqnarray}

The candidates of class 3 states are
\begin{eqnarray}
\lb (0,2),\,(0,3)\rb~,~~~ \lb (0,1),\,(0,4) \rb~.
\end{eqnarray}
The first one is again reduced to a class 2 state because of 
the character identity
\begin{eqnarray}
 \chim^{}(0,2;\tau,z)+\chim^{}(0,3;\tau,z)= \chi(p=0,j=2;\tau,z) 
    (\equiv  \Ch{(\sNS)}{}\left(h=1,|Q|=2;\tau,z\right))~.
\end{eqnarray}
The self-overlap of the second one is calculated as
\begin{eqnarray}
&&e^{\pi \frac{5z^2}{T}} \, \bra{B;(0,1),\,(0,4)}
e^{-\pi T H^{(c)}} e^{i\pi z (J_0+\tJ_0)} 
\ket{B;(0,1),\,(0,4)} 
= \int_0^{\infty}dp\,
\sum_{j=0,\pm 1, 2}\, \rho_{j}(p)\chi(p,j;it,z')~, \nn
&& \rho_0(p) = 2\delta(p) + 4 \int_0^{\infty}dp'\, 
\left(\coth(\pi p')\coth(2\pi p')-1\right) \cos(2\pi p p')~, \nn
&& \rho_1(p)=\rho_{-1}(p)= 2\int_0^{\infty}dp'\,
\frac{\cos(2\pi p p')}{\sinh(\pi p')\sinh(2\pi p')} \nn
&& \rho_2(p)=2\delta(p)~.
\label{c hat 5 class 3 amplitude}
\end{eqnarray}
The spectral densities are indeed positive, 
and hence $\ket{B;(0,1),\,(0,4)}$ are the Cardy state.

~



\newpage
\begin{thebibliography}{99}



\bibitem{KutS}
D.~Kutasov and N.~Seiberg,
Phys.\ Lett.\ B {\bf 251}, 67 (1990);
D.~Kutasov,
{\it ``Some properties of (non)critical strings,''}
Lectures given at ICTP Spring School on String Theory and Quantum Gravity, Trieste, Italy, Apr 15-23, 1991,
arXiv:hep-th/9110041.



\bibitem{FZZ}
V.~Fateev, A.~B.~Zamolodchikov and A.~B.~Zamolodchikov,
arXiv:hep-th/0001012.

\bibitem{Teschner}
J.~Teschner,
arXiv:hep-th/0009138.

\bibitem{ZZ}
A.~B.~Zamolodchikov and A.~B.~Zamolodchikov,
arXiv:hep-th/0101152.



\bibitem{FH}
T.~Fukuda and K.~Hosomichi,
Nucl.\ Phys.\ B {\bf 635}, 215 (2002)
[arXiv:hep-th/0202032].

\bibitem{ARS}
C.~Ahn, C.~Rim and M.~Stanishkov,
Nucl.\ Phys.\ B {\bf 636}, 497 (2002)
[arXiv:hep-th/0202043].



\bibitem{GV}
D.~Ghoshal and C.~Vafa,
Nucl.\ Phys.\ B {\bf 453}, 121 (1995)
[arXiv:hep-th/9506122].


\bibitem{OV}
H.~Ooguri and C.~Vafa,
Nucl.\ Phys.\ B {\bf 463}, 55 (1996)
[arXiv:hep-th/9511164].


\bibitem{ABKS}
O.~Aharony, M.~Berkooz, D.~Kutasov and N.~Seiberg,
JHEP {\bf 9810}, 004 (1998)
[arXiv:hep-th/9808149].


\bibitem{GKP}
A.~Giveon, D.~Kutasov and O.~Pelc,
JHEP {\bf 9910}, 035 (1999)
[arXiv:hep-th/9907178].


\bibitem{GK}
A.~Giveon and D.~Kutasov,
JHEP {\bf 9910}, 034 (1999)
[arXiv:hep-th/9909110];
A.~Giveon and D.~Kutasov,
JHEP {\bf 0001}, 023 (2000)
[arXiv:hep-th/9911039].


\bibitem{Pelc}
O.~Pelc,
JHEP {\bf 0003}, 012 (2000)
[arXiv:hep-th/0001054].


\bibitem{ES1}
T.~Eguchi and Y.~Sugawara,
Nucl.\ Phys.\ B {\bf 577}, 3 (2000)
[arXiv:hep-th/0002100].


\bibitem{Mizoguchi}
S.~Mizoguchi,
JHEP {\bf 0004}, 014 (2000)
[arXiv:hep-th/0003053].


\bibitem{Yamaguchi}
S.~Yamaguchi,
Nucl.\ Phys.\ B {\bf 594}, 190 (2001)
[arXiv:hep-th/0007069];
Phys.\ Lett.\ B {\bf 509}, 346 (2001)
[arXiv:hep-th/0102176];
JHEP {\bf 0201}, 023 (2002)
[arXiv:hep-th/0112004].


\bibitem{NN}
M.~Naka and M.~Nozaki,
Nucl.\ Phys.\ B {\bf 599}, 334 (2001)
[arXiv:hep-th/0010002].

\bibitem{HK2}
K.~Hori and A.~Kapustin,
JHEP {\bf 0211}, 038 (2002)
[arXiv:hep-th/0203147].

\bibitem{Lerche}
W.~Lerche,
arXiv:hep-th/0006100.


\bibitem{LLS}
W.~Lerche, C.~A.~Lutken and C.~Schweigert,
Nucl.\ Phys.\ B {\bf 622}, 269 (2002)
[arXiv:hep-th/0006247].


\bibitem{ES2}
T.~Eguchi and Y.~Sugawara,
Nucl.\ Phys.\ B {\bf 598}, 467 (2001)
[arXiv:hep-th/0011148].

\bibitem{MV}
J.~McGreevy and H.~Verlinde,
arXiv:hep-th/0304224.

\bibitem{KMS}
I.~R.~Klebanov, J.~Maldacena and N.~Seiberg,
JHEP {\bf 0307}, 045 (2003)
[arXiv:hep-th/0305159].

\bibitem{MTV}
J.~McGreevy, J.~Teschner and H.~Verlinde,
arXiv:hep-th/0305194.

\bibitem{Martinec}
E.~J.~Martinec,
arXiv:hep-th/0305148.

\bibitem{AKK}
S.~Y.~Alexandrov, V.~A.~Kazakov and D.~Kutasov,
JHEP {\bf 0309}, 057 (2003)
[arXiv:hep-th/0306177].


\bibitem{TT}
T.~Takayanagi and N.~Toumbas,
JHEP {\bf 0307}, 064 (2003)
[arXiv:hep-th/0307083].


\bibitem{DKKMMS}
M.~R.~Douglas, I.~R.~Klebanov, D.~Kutasov, J.~Maldacena, E.~Martinec and N.~Seiberg,
arXiv:hep-th/0307195.

\bibitem{GIR}
D.~Gaiotto, N.~Itzhaki and L.~Rastelli,
arXiv:hep-th/0307221.


\bibitem{GutK}
M.~Gutperle and P.~Kraus,
arXiv:hep-th/0308047.


\bibitem{Sen}
A.~Sen,
arXiv:hep-th/0308068.


\bibitem{MMV}
J.~McGreevy, S.~Murthy and H.~Verlinde,
arXiv:hep-th/0308105.

\bibitem{Kapustin}
A.~Kapustin,
arXiv:hep-th/0308119.


\bibitem{GKPS}
A.~Giveon, A.~Konechny, A.~Pakman and A.~Sever,
arXiv:hep-th/0309056.

\bibitem{KStro}
J.~L.~Karczmarek and A.~Strominger,
arXiv:hep-th/0309138.

\bibitem{DRSVW}
O.~DeWolfe, R.~Roiban, M.~Spradlin, A.~Volovich and J.~Walcher,
arXiv:hep-th/0309148.


\bibitem{KMS2}
I.~R.~Klebanov, J.~Maldacena and N.~Seiberg,
arXiv:hep-th/0309168.


\bibitem{DD}
S.~Dasgupta and T.~Dasgupta,
arXiv:hep-th/0310106.


\bibitem{Alex}
S.~Alexandrov,
arXiv:hep-th/0310135.


\bibitem{GKap}
J.~Gomis and A.~Kapustin,
arXiv:hep-th/0310195.


\bibitem{Mukhi}
S.~Mukhi,
arXiv:hep-th/0310287.


\bibitem{BH}
O.~Bergman and S.~Hirano,
arXiv:hep-th/0311068.






\bibitem{BFK}
W.~Boucher, D.~Friedan and A.~Kent,
Phys.\ Lett.\ B {\bf 172}, 316 (1986).


\bibitem{Dobrev}
V.~K.~Dobrev,
Phys.\ Lett.\ B {\bf 186}, 43 (1987).



\bibitem{ET}
T.~Eguchi and A.~Taormina,
Phys.\ Lett.\ B {\bf 200}, 315 (1988);
Phys.\ Lett.\ B {\bf 210}, 125 (1988).



\bibitem{EOTY}
T.~Eguchi, H.~Ooguri, A.~Taormina and S.~K.~Yang,
Nucl.\ Phys.\ B {\bf 315}, 193 (1989).



\bibitem{Odake}
S.~Odake,
Mod.\ Phys.\ Lett.\ A {\bf 4}, 557 (1989);
Int.\ J.\ Mod.\ Phys.\ A {\bf 5}, 897 (1990).


\bibitem{HS}
Y.~Hikida and Y.~Sugawara,
JHEP {\bf 0210}, 067 (2002)
[arXiv:hep-th/0207124].





\bibitem{Miki}
K.~Miki,
Int.\ J.\ Mod.\ Phys.\ A {\bf 5}, 1293 (1990).

\bibitem{Ishibashi}
N.~Ishibashi,
Mod.\ Phys.\ Lett.\ A {\bf 4}, 251 (1989).

\bibitem{Cardy}
J.~L.~Cardy,
Nucl.\ Phys.\ B {\bf 324}, 581 (1989).






\bibitem{OOY}
H.~Ooguri, Y.~Oz and Z.~Yin,
Nucl.\ Phys.\ B {\bf 477}, 407 (1996)
[arXiv:hep-th/9606112].






\bibitem{BF}
P.~Baseilhac and V.~A.~Fateev,
Nucl.\ Phys.\ B {\bf 532}, 567 (1998)
[arXiv:hep-th/9906010].

\bibitem{AKRS}
C.~Ahn, C.~Kim, C.~Rim and M.~Stanishkov,
arXiv:hep-th/0210208.

\bibitem{Teschner3}
J.~Teschner,
Phys.\ Lett.\ B {\bf 363}, 65 (1995)
[arXiv:hep-th/9507109].


\bibitem{Seiberg-L}
N.~Seiberg,
Prog.\ Theor.\ Phys.\ Suppl.\  {\bf 102}, 319 (1990).


\bibitem{AY}
C.~Ahn and M.~Yamamoto,
arXiv:hep-th/0310046.






\bibitem{RS}
A.~Recknagel and V.~Schomerus,
Nucl.\ Phys.\ B {\bf 531}, 185 (1998)
[arXiv:hep-th/9712186].


\bibitem{HNS}
Y.~Hikida, M.~Nozaki and Y.~Sugawara,
Nucl.\ Phys.\ B {\bf 617}, 117 (2001)
[arXiv:hep-th/0101211].



\bibitem{MSY}
A.~Maloney, A.~Strominger and X.~Yin,
arXiv:hep-th/0302146.



\bibitem{GS1}
M.~Gutperle and A.~Strominger,
JHEP {\bf 0204}, 018 (2002)
[arXiv:hep-th/0202210].


\bibitem{Sen-RT}
A.~Sen,
JHEP {\bf 0204}, 048 (2002)
[arXiv:hep-th/0203211];
JHEP {\bf 0207}, 065 (2002)
[arXiv:hep-th/0203265];
Mod.\ Phys.\ Lett.\ A {\bf 17}, 1797 (2002)
[arXiv:hep-th/0204143].


\bibitem{Strominger}
A.~Strominger,
arXiv:hep-th/0209090.


\bibitem{LLM}
N.~Lambert, H.~Liu and J.~Maldacena,
arXiv:hep-th/0303139.




\bibitem{Sen-review}
A.~Sen,
arXiv:hep-th/9904207.



\bibitem{Sugawara}
Y.~Sugawara,
JHEP {\bf 0308}, 008 (2003)
[arXiv:hep-th/0307034].


\bibitem{AW}
J. J. Atick and E. Witten,
Nucl. Phys. B {\bf 310}, 291 (1988).


\bibitem{MP}
E.~Marinari and G.~Parisi,
Phys.\ Lett.\ B {\bf 240}, 375 (1990).


\bibitem{DiFK}
P.~Di Francesco and D.~Kutasov,
Nucl.\ Phys.\ B {\bf 375}, 119 (1992)
[arXiv:hep-th/9109005].



\bibitem{KKK}
V.~Kazakov, I.~K.~Kostov and D.~Kutasov,
Nucl.\ Phys.\ B {\bf 622}, 141 (2002)
[arXiv:hep-th/0101011].


\bibitem{LNT}
F.~Larsen, A.~Naqvi and S.~Terashima,
JHEP {\bf 0302}, 039 (2003)
[arXiv:hep-th/0212248].


\bibitem{HK1}
K.~Hori and A.~Kapustin,
JHEP {\bf 0108}, 045 (2001)
[arXiv:hep-th/0104202].


\bibitem{FZZ2}
V.~Fateev, A.~B.~Zamolodchikov and A.~B.~Zamolodchikov,
unpublished, as cited in \cite{KKK}.



\bibitem{GS2}
M.~Gutperle and A.~Strominger,
Phys.\ Rev.\ D {\bf 67}, 126002 (2003)
[arXiv:hep-th/0301038].


\bibitem{ST}
A.~Strominger and T.~Takayanagi,
Adv.\ Theor.\ Math.\ Phys.\  {\bf 7}, 2 (2003)
[arXiv:hep-th/0303221].


\bibitem{Schomerus}
V.~Schomerus,
arXiv:hep-th/0306026.










\bibitem{RibS}
S.~Ribault and V.~Schomerus,
arXiv:hep-th/0310024.


\bibitem{PST}
B.~Ponsot, V.~Schomerus and J.~Teschner,
JHEP {\bf 0202}, 016 (2002)
[arXiv:hep-th/0112198].


\bibitem{GKSch}
A.~Giveon, D.~Kutasov and A.~Schwimmer,
Nucl.\ Phys.\ B {\bf 615}, 133 (2001)
[arXiv:hep-th/0106005].


\bibitem{LOP}
P.~Lee, H.~Ooguri and J.~w.~Park,
Nucl.\ Phys.\ B {\bf 632}, 283 (2002)
[arXiv:hep-th/0112188].


\bibitem{STT}
A.~M.~Semikhatov, I.~Y.~Tipunin and A.~Taormina,
arXiv:math.qa/0311314.
















\end{thebibliography}
\end{document}